\documentclass[12pt, a4paper]{article}

\RequirePackage[l2tabu, orthodox]{nag}

\usepackage[top=20mm,bottom=20mm,left=25mm,right=25mm]{geometry}

\usepackage{amsmath,amssymb,amsfonts}

\usepackage{centernot}

\usepackage{underbracket}

\usepackage[nottoc]{tocbibind}

\usepackage{pgf}
\usepackage{tikz}

\usepackage{multirow}

\usepackage{pgf}
\usepackage{tikz}

\usepackage{xcolor}

\newcommand*{\colorboxed}{}
\def\colorboxed#1#{%
  \colorboxedAux{#1}%
}
\newcommand*{\colorboxedAux}[3]{%
  \begingroup
    \colorlet{cb@saved}{.}%
    \color#1{#2}%
    \boxed{%
      \color{cb@saved}%
      #3%
    }%
  \endgroup
}

\usepackage[most]{tcolorbox}

\tcbset{
    frame code={}
    center title,
    left=0pt,
    right=0pt,
    top=0pt,
    bottom=0pt,
    colback=green!15,
    colframe=white,
    enlarge left by=0mm,
    arc=0pt,outer arc=0pt,
    breakable,
    }

\usepackage{amsthm}

\usepackage{thmtools}

\makeatletter
\def\@endtheorem{\endtrivlist}
\makeatother

\newtheoremstyle{break}{9pt}{9pt}{\itshape}{}{\bfseries}{}{\newline}{}
\theoremstyle{break}    

\newtheorem{exo}{Exercise}[section]
\newtheorem{hyp}[exo]{Axiom}

\newtheorem{defn}[exo]{Definition}

\usepackage[colorlinks=true,linktoc=all,linkcolor=black,citecolor=red,urlcolor=blue]{hyperref}

\title{\bfseries Minimal lectures on two-dimensional \\ conformal field theory}

\author{Sylvain Ribault \vspace{2mm}
\\
{\normalsize CEA Saclay, Institut de Physique Th\'eorique}
 \\
 {\footnotesize \ttfamily sylvain.ribault@ipht.fr }
}

\begin{document}

\maketitle

\begin{abstract}
We provide a brief but self-contained review of two-dimensional conformal field theory, from the basic principles to some of the simplest models. From the representations of the Virasoro algebra on the one hand, and the state-field correspondence on the other hand, we deduce Ward identities and Belavin--Polyakov--Zamolodchikov equations for correlation functions. We then explain the principles of the conformal bootstrap method, and introduce conformal blocks. This allows us to define and solve minimal models and Liouville theory. In particular, we study their
three- and four-point functions, and discuss their existence and uniqueness. In appendices, we introduce the free boson theory (with an arbitrary central charge), and the modular bootstrap in minimal models.
\end{abstract}

\vspace{5mm}

 \textit{Based on lectures given at the school on ``Quantum integrable systems, conformal field theories and stochastic processes'' (Carg\`ese, September 2016), and at the ``Young Researchers Integrability School'' (Vienna, February 2019).}
\vspace{2mm}

\textit{
An earlier version of this text was published in SciPost Physics Lecture Notes: arXiv's fourth version} \href{https://arxiv.org/abs/1609.09523v4}{\texttt{arXiv:1609.09523v4}}. 

\clearpage

\tableofcontents

\hypersetup{linkcolor=blue}

\numberwithin{equation}{section}
\setcounter{section}{-1}

\section{Introduction}

Two-dimensional CFTs belong to the rare cases of quantum field theories that can be exactly solved, thanks to their infinite-dimensional symmetry algebras. They are interesting for their applications to statistical physics, they are the technical basis of string theory in the worldsheet approach, and they can guide the exploration of higher-dimensional CFTs. 

We will introduce the main ideas of two-dimensional CFT in the conformal bootstrap approach, and focus on the simplest nontrivial models that have been solved: minimal models and Liouville theory. Rather than following the history of the subject, we try to derive the results in the simplest possible way. 
While not claiming mathematical rigour, we explicitly state the axioms that underlie our derivations.
This is supposed to facilitate generalizations, for example to CFTs based on larger symmetry algebras, or to non-diagonal CFTs \cite{mr17}.

Our first axioms will specify how the Virasoro symmetry algebra acts on fields, and the existence and properties of the operator product expansion. 
Next, we introduce additional axioms that single out either minimal models, or Liouville theory.
We will then check that these theories actually exist, by studying their four-point functions.
It is the success of such checks, more than a priori considerations, that justifies our choice of axioms. 

Our main tool for solving CFTs is crossing symmetry of the sphere four-point function. We will however introduce two other tools as side subjects:
\begin{itemize}
 \item The free boson (Appendix \ref{sec:fb}) is not needed in our approach, because we do not build CFTs as perturbed free theories, as is done in the Lagrangian approach. However, it is a good preparation to the study of CFTs based on larger symmetry algebras, in particular WZW models. 
 \item The modular bootstrap (Appendix \ref{sec:mb}) focuses on torus partition functions: less interesting than sphere four-point functions, but also much simpler, so they can be tractable even in complicated models. 
\end{itemize}
This text  aims to be self-contained, except at the very end when we will refer to \cite{zz90} for the properties of generic conformal blocks.
For a more detailed text in the same spirit, see the review article \cite{rib14}. For a wider and more advanced review, and a guide to the recent literature, see Teschner's text \cite{tes17}. 
The Bible of rational conformal field theory is of course the epic textbook \cite{fms97}. And Cardy's lecture notes \cite{car08} provide an introduction to the statistical physics applications of conformal field theory.

\begin{exo}[Update Wikipedia]
 ~\label{exo:wiki}
 How would you rate Wikipedia's coverage of two-dimensional CFT? For a list of some relevant articles, see 
 \href{https://en.wikipedia.org/wiki/User:Sylvain_Ribault/YRIS2019}{this page}.
 Correct and update these articles when needed. Are there other relevant articles? Which articles should be created?
\end{exo}

\subsection*{Acknowledgements}

I am grateful to the organizers of the Carg\`ese school, for challenging me to explain Liouville theory in about four hours. I am grateful to the organizers of the Vienna school, for the opportunity to fit these lectures into a school on various aspects of two-dimensional CFT. 

I wish to thank Bertrand Eynard, Riccardo Guida, Yifei He, and Andr\'e Voros for helpful suggestions and comments.
I am grateful to the participants of the Carg\`ese and Vienna schools, for their stimulating participation in the lectures.
I wish to thank the SciPost editor and reviewers for their feedback and \href{https://scipost.org/submission/1609.09523v2/}{suggestions}, which led to many improvements, both perturbative and non-perturbative.


\section{The Virasoro algebra and its representations}

\subsection{Algebra}

By definition, conformal transformations are transformations that preserve angles. 
In two dimensions with a complex coordinate $z$, any holomorphic transformation preserves angles.
Infinitesimal conformal transformations are holomorphic functions close to the identity function, 
\begin{align}
 z \mapsto z + \epsilon z^{n+1}\qquad (n\in\mathbb{Z}\ , \ \epsilon\ll 1) \ .
\end{align}
These transformations act on functions of $z$ via the differential operators 
\begin{align}
 \ell_n = -z^{n+1}\frac{\partial}{\partial z}\ ,
\end{align}
and these operators generate the Witt algebra, with commutation relations
\begin{align}
 [\ell_n,\ell_m ] = (n-m)\ell_{m+n}\ .
\end{align}

The generators $(\ell_{-1},\ell_0,\ell_1)$ generate a subalgebra called the algebra of infinitesimal global conformal transformations and isomorphic to $s\ell_2$.  The corresponding Lie group is the group of global conformal transformations of 
the Riemann sphere $\mathbb{C}\cup \{\infty\}$,
\begin{align}
 z \mapsto \frac{az+b}{cz+d}\quad , \quad (a,b,c,d\in \mathbb{C},\ ad-bc\neq 0)\ .
\end{align}


\begin{exo}[Global conformal group of the sphere]
 ~\label{exo:sphere}
Show that the global conformal group of the sphere is $PSL_2(\mathbb{C})$, and includes translations, rotations, and dilations. 
\end{exo}

In a quantum theory, symmetry transformations act projectively on states. 
Projective representations of an algebra are equivalent to representations of a centrally extended algebra. 
This is why we always look for central extensions of symmetry algebras.

\begin{defn}[Virasoro algebra]
 ~\label{def:vir}
 The central extension of the Witt algebra is called the Virasoro algebra. It has the generators $(L_n)_{n\in\mathbb{Z}}$ and $\mathbf 1$, and the commutation relations
 \begin{align}
  [\mathbf 1, L_n] = 0 \quad , \quad [L_n,L_m] = (n-m)L_{n+m} +\frac{c}{12}(n-1)n(n+1)\delta_{n+m,0}\mathbf 1 \ ,
  \label{eq:vir}
 \end{align}
 where the number $c$ is called the central charge. (The notation $c\mathbf 1$ stands for a central generator that always has the same eigenvalue $c$ within a given conformal field theory.)
\end{defn}

\begin{exo}[Uniqueness of the Virasoro algebra]
 ~\label{exo:vir}
 Show that the Virasoro algebra is the unique central extension of the Witt algebra.
\end{exo}

\subsection{Representations}

The spectrum, i.e. the space of states, must be a representation of the Virasoro algebra. Let us now make assumptions on what type of representation it can be.

\begin{hyp}[Representations that can appear in the spectrum]
 ~\label{hyp:rep}
 The spectrum is a direct sum of irreducible representations. In the spectrum, $L_0$ is diagonalizable, and the real part of its eigenvalues is bounded from below.
\end{hyp}
Why this special role for $L_0$? Because we want to interpret it as the energy operator. 
We however do not assume that $L_0$ eigenvalues are real or that the spectrum is a Hilbert space: this would restrict the central charge to be real. The $L_0$ eigenvalue of an $L_0$ eigenvector is called its conformal dimension. The action of $L_n$ shifts conformal dimensions by $-n$:
\begin{align}
 L_0|v\rangle = \Delta|v\rangle \quad \Rightarrow\quad  L_0 L_n|v\rangle = L_nL_0|v\rangle + [L_0, L_n] |v\rangle  = (\Delta-n)L_n|v\rangle \ .
\end{align}
Let us consider an irreducible representation that is allowed by our axiom. In this representation, all $L_0$ eigenvalues differ by integers, and there is an eigenvector $|v\rangle$ whose eigenvalue $\Delta$ is smallest in real part. 
If follows that $L_n|v\rangle =0$ for $n>0$, and $|v\rangle $ is called a primary state.

\begin{defn}[Primary and descendent states, level, Verma module]
 ~\label{def:prim}
 A primary state with conformal dimension $\Delta$ is a state $|v\rangle\neq 0$ such that 
 \begin{align}
  L_0 |v\rangle = \Delta |v\rangle \quad , \quad L_{n>0} |v\rangle = 0\ .
 \end{align}
The Verma module $\mathcal V_\Delta$ is the representation whose basis is 
 $
 \left\{ \prod_{i=1}^k L_{-n_i} |v\rangle\right\}_{ 0<n_1\leq \dots \leq n_k}
 $.
The state $\prod_{i=1}^k L_{-n_i} |v\rangle $ has the conformal dimension $\Delta+N$, where $N=\sum_{i=1}^k n_i\geq 0$ is called the level. A state of level $N\geq 1$ is called a descendent state.
\end{defn}
Let us plot a basis of primary and descendent states up to the level $3$:
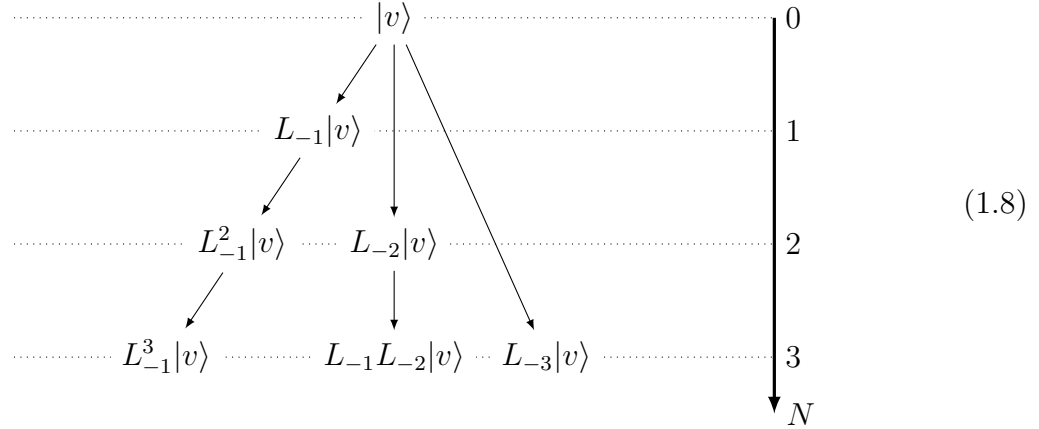
\begin{align}
 \begin{tikzpicture}[scale = .25, baseline=(current  bounding  box.center)]
  \draw[-latex, very thick] (20, 0) -- (20, -21) node [right] {$N$};
  \foreach \x in {0, ..., 3}
  {
  \draw [dotted] (-20, {-6*\x}) -- (20, {-6*\x}) node [right] {${\x}$};
  }
  \node[fill = white] at (0, 0) (0) {$|v\rangle$};
  \node[fill = white] at (-4,-6) (1) {$L_{-1}|v\rangle$};
  \node[fill = white] at (-8, -12) (11) {$L_{-1}^2|v\rangle$};
  \node[fill = white] at (-12, -18) (111) {$L_{-1}^3|v\rangle$};
  \node[fill = white] at (0,-12) (2) {$L_{-2}|v\rangle$};
  \node[fill = white] at (0,-18) (12) {$L_{-1}L_{-2}|v\rangle$};
  \node[fill = white] at (8,-18) (3) {$L_{-3}|v\rangle$};
  \draw[-latex] (0) -- (1);
  \draw[-latex] (1) -- (11);
  \draw[-latex] (11) -- (111);
  \draw[-latex] (0) -- (2);
  \draw[-latex] (0) -- (3);
  \draw[-latex] (2) -- (12);
 \end{tikzpicture}
\end{align}
We need not include the state $L_{-2}L_{-1}|v\rangle$, due to $L_{-2}L_{-1} = L_{-1}L_{-2} - L_{-3}$.

Are Verma modules reducible representations? i.e. do they have nontrivial subrepresentations? In any subrepresentation of a Verma module, $L_0$ is again diagonalizable and bounded from below, so there must be a primary state $|\chi\rangle$. If the subrepresentation differs from the Verma module, that primary state must differ from $|v\rangle$, and therefore be a descendent of $|v\rangle$.

\subsection{Null vectors and degenerate representations}\label{sec:nv}

\begin{defn}[Null vectors]
 ~\label{def:nv}
 A descendent state that is also primary is called a null vector or singular vector.
\end{defn}
In the Verma module $\mathcal V_\Delta$, let us look for null vectors at the level $N=1$. For $n\geq 1$ we have 
\begin{align}
L_n L_{-1}|v\rangle = [L_n, L_{-1}] |v\rangle = (n+1) L_{n-1}|v\rangle = 
\left\{\begin{array}{ll} 0 &  \quad \text{if } n\geq 2\ , \\ 2\Delta |v\rangle & \quad \text{if } n = 1\ . \end{array}\right. 
\end{align}
So $L_{-1}|v\rangle$ is a null vector if and only if $\Delta=0$, and the Verma module $\mathcal V_0$ is reducible.
Let us now look for null vectors at the level $N=2$. Let $|\chi\rangle = (L_{-1}^2 + a L_{-2})|v\rangle$, then $L_{n\geq 3} |\chi \rangle =0$. 

\begin{exo}[Level two null vectors]
 ~\label{exo:level2}
 Compute  $L_1|\chi\rangle$ and $L_2|\chi\rangle$, and find 
 \begin{align}
  L_1 |\chi\rangle = \left((4\Delta+2) + 3a\right) L_{-1}|v\rangle
  \quad , \quad L_2 |\chi\rangle= \left(6\Delta + (4\Delta +\tfrac12 c)a\right) |v\rangle\ .
 \end{align}
 Requiring that $L_1|\chi\rangle$ and $L_2|\chi\rangle$ vanish, find the coefficient $a$, and show that
 \begin{align}
 \Delta = \frac{1}{16}\left( 5-c\pm\sqrt{(c-1)(c-25)} \right) \ .
 \label{eq:dpm}
\end{align}
\end{exo}
In order to simplify this formula, let us introduce other notations for $c$ and $\Delta$. We define
\begin{align}
 \text{the background charge } Q \ , & \quad c = 1+6Q^2\ , \quad \text{up to } Q \mapsto -Q\ ,
 \label{eq:cqb}
 \\
 \text{the coupling constant } b \ , & \quad Q = b+\frac{1}{b} \ , \quad \text{up to } b\mapsto \pm b^{\pm 1}\ ,
 \\
 \text{the momentum } P\  , &\quad \Delta = \frac{Q^2}{4}-P^2\ , \quad \text{up to reflections } P \mapsto -P\ .
\label{eq:refm}
 \end{align}
The condition \eqref{eq:dpm} for the existence of a level two null vector becomes 
\begin{align}
 P = \frac12\left( b+ b^{-1} + b^{\pm 1}\right)\ .
\end{align}
Let us summarize the momentums of the Verma modules that have null vectors at levels $N=1,2$, and the null vectors themselves:
\begin{align}
\renewcommand{\arraystretch}{1.3}
\begin{array}{|c|c|c|c|}
\hline 
N & \langle r,s\rangle & P_{\langle r,s\rangle} &  L_{\langle r,s\rangle} 
\\
\hline\hline
1 & \langle 1,1\rangle & \frac12\left(b+b^{-1}\right) &  L_{-1}
\\
\hline
\multirow{2}{*}{2} & 
\langle 2,1\rangle & \frac12\left( 2b+ b^{-1}\right)  & L_{-1}^2 + b^2 L_{-2}
\\
\cline{2-4}
& \langle 1,2\rangle & \frac12\left( b+ 2b^{-1} \right) & L_{-1}^2 + b^{-2} L_{-2} 
\\
\hline
rs & \langle r,s\rangle &  \frac12\left(rb+sb^{-1}\right) & L_{-1}^{rs} + \cdots 
\\
\hline
\end{array}
\label{eq:ars}
\end{align}
The generalization to higher levels $N\geq 3$ is that the dimensions of Verma modules with null vectors are labelled by positive integers $r,s$ such that $N=rs$. We write these dimensions $\Delta_{\langle r,s\rangle}$, and the corresponding momentums $P_{\langle r,s\rangle}$. We accept these results for now, see the later Exercise \ref{exo:hdr} for a derivation.

If $\Delta\notin\{\Delta_{\langle r,s\rangle}\}_{r,s\in\mathbb{N}^*}$, then $\mathcal V_\Delta$ is irreducible. If $\Delta = \Delta_{\langle r,s\rangle}$, then $\mathcal V_\Delta$ contains a nontrivial submodule, generated by the null vector and its descendent states. For generic values of the central charge $c$, this submodule is the Verma module $\mathcal V_{\Delta_{\langle r,s\rangle}+rs}$.

\begin{defn}[Degenerate representation]
 ~\label{def:deg}
The coset of the reducible Verma module $\mathcal V_{\Delta_{\langle r,s\rangle}}$ by its Verma submodule $\mathcal V_{\Delta_{\langle r,s\rangle}+rs}$ is an irreducible module $\mathcal{R}_{\langle r,s\rangle}$, which is called a degenerate representation:
\begin{align}
 \mathcal{R}_{\langle r,s\rangle} = \frac{\mathcal V_{\Delta_{\langle r,s\rangle}}}{\mathcal V_{\Delta_{\langle r,s\rangle}+rs}}\ .
\end{align}
In this representation, the null vector vanishes,
\begin{align}
 L_{\langle r,s\rangle}|v\rangle = 0\ .
\end{align}
\end{defn}

\section{Fields and correlation functions}\label{sec:cft}

Now that we understand the algebraic structure of conformal symmetry in two dimensions, let us study how the Virasoro algebra acts on objects that live on the Riemann sphere -- the fields of conformal field theory. We will not try to construct the fields, or to specify the space they live in: it is enough to view fields as notations for describing the properties of correlation functions, and to understand equations for fields as valid inside correlation functions.

\subsection{Fields}

\begin{hyp}[State-field correspondence]
 ~\label{hyp:sfc}
For any state $|w\rangle$ in the spectrum, there is an associated field $V_{|w\rangle}(z)$. The map $|w\rangle \mapsto V_{|w\rangle}(z)$ is linear and injective. We define the action of the Virasoro algebra on such fields as 
\begin{align}
 L_n V_{|w\rangle}(z) =   V_{L_n|w\rangle}(z)\ .
\end{align}
We also sometimes use the notation $L_n^{(z)} V_{|w\rangle}(z)=L_n V_{|w\rangle}(z)$. 
\end{hyp}

\begin{defn}[Primary field, descendent field, degenerate field]
~\label{def:pfdf}
Let $|v\rangle$ be the primary state of the Verma module $\mathcal V_\Delta$.
We define the primary field $V_\Delta(z)=V_{|v\rangle}(z)$. This field obeys
\begin{align}
 L_0 V_\Delta(z) = \Delta V_\Delta(z) \quad , \quad  L_{n> 0} V_\Delta(z) = 0 \ .
\end{align}
Similarly, descendent fields correspond to descendent states. And the degenerate field $V_{\langle r,s\rangle}(z)$ corresponds to the primary state of the degenerate representation $\mathcal{R}_{\langle r,s\rangle}$, and therefore obeys 
\begin{align}
L_0 V_{\langle r,s\rangle}(z) = \Delta_{\langle r,s\rangle} V_{\langle r,s\rangle}(z) \quad , \quad  L_{n> 0} V_{\langle r,s\rangle}(z) = 0 \quad , \quad L_{\langle r, s\rangle} V_{\langle r,s\rangle}(z) = 0\ .
\end{align}
\end{defn}

\begin{hyp}[Dependence of fields on $z$]
 ~\label{hyp:geom}
 For any field $V(z)$, we have 
 \begin{align}
  \frac{\partial}{\partial z} V(z) = L_{-1} V(z)  \ .
  \label{pvlv}
 \end{align}
\end{hyp}
Using this axiom for both $V(z)$ and $L_n^{(z)}V(z)$, we find how $L_n^{(z)}$ depends on $z$:
\begin{align}
 \frac{\partial}{\partial z} L_n^{(z)} = [L_{-1}^{(z)},L_n^{(z)}]= -(n+1)L_{n-1}^{(z)}\ ,\qquad (\forall n\in\mathbb{Z})\ .
\end{align}
These infinitely many equations can be encoded into one functional equation,
\begin{align}
 \frac{\partial}{\partial z} \sum_{n\in\mathbb{Z}} \frac{L_n^{(z)}}{(y-z)^{n+2}} = 0\ .
\end{align}

\begin{defn}[Energy-momentum tensor]
 ~\label{def:em}
 The energy-momentum tensor is a field, that we define by the formal Laurent series
 \begin{align}
  T(y) = \sum_{n\in\mathbb{Z}} \frac{L_n^{(z)}}{(y-z)^{n+2}} \ .
 \end{align}
In other words, for any field $V(z)$, we have 
\begin{align}
 T(y)V(z) = \sum_{n\in\mathbb{Z}} \frac{L_n V(z)}{(y-z)^{n+2}}\quad , \quad L_n V(z) = \frac{1}{2\pi i} \oint_{z}dy\ (y-z)^{n+1} T(y)V(z)\ .
 \label{eq:lvtv}
\end{align}
\end{defn}
In the case of a primary field $V_\Delta(z)$, using eq. \eqref{pvlv} and writing regular terms as $O(1)$, this definition reduces to
\begin{align}
 T(y)V_\Delta(z) \underset{y\to z}{=} \frac{\Delta}{(y-z)^2} V_\Delta(z) + \frac{1}{y-z} \frac{\partial}{\partial z} V_\Delta(z) + O(1)\ . 
 \label{eq:tvd}
\end{align}
This is our first example of an operator product expansion.

The energy-momentum tensor $T(y)$ is locally holomorphic as a function of $y$, and acquires poles in the presence of other fields. Since we are on the Riemann sphere, it must also be holomorphic at $y=\infty$. 

\begin{hyp}[Behaviour of $T(y)$ at infinity]
~\label{hyp:ti}
 \begin{align}
 T(y) \underset{y\to\infty} = O\left(\frac{1}{y^4}\right)\ .
 \label{eq:tinf}
\end{align}
\end{hyp}

\subsection{Correlation functions and Ward identities}

\begin{defn}[Correlation function]
~\label{def:cor}
 To $N$ fields $V_1(z_1), \dots ,V_N(z_N)$ with $i\neq j\implies z_i\neq z_j$, we associate a number called their correlation function or $N$-point function, and denoted as 
 \begin{align}
  \Big< V_1(z_1) \cdots V_N(z_N) \Big>\ .
 \end{align}
For example, $\left< \prod_{i=1}^N V_{\Delta_i}(z_i) \right>$ is a function of $\{z_i\}, \{\Delta_i\}$ and $c$.
Correlation functions depend linearly on fields, and in particular $\frac{\partial}{\partial z_1} \left< V_1(z_1) \cdots V_N(z_N) \right> = \left< \frac{\partial}{\partial z_1}V_1(z_1) \cdots V_N(z_N) \right>$.
\end{defn}

\begin{hyp}[Commutativity of fields]
 ~\label{hyp:ass}
 Correlation functions do not depend on the order of the fields,
 \begin{align}
  V_1(z_1) V_2(z_2) = V_2(z_2)V_1(z_1)\ .
 \end{align}
\end{hyp}

\begin{exo}[Virasoro algebra and OPE] 
~\label{exott}
Show that the commutation relations \eqref{eq:vir} of the Virasoro algebra are equivalent to the following OPE of the field $T(y)$ with itself,
\begin{align}
 T(y)T(z) \underset{y\to z}{=} \frac{\frac{c}{2}}{(y-z)^4} + \frac{2T(z)}{(y-z)^2} + \frac{\partial T(z)}{y-z} + O(1)\ .
\label{tt}
\end{align}
\end{exo}

Let us work out the consequences of conformal symmetry for correlation functions.
In order to study an $N$-point function $Z$ of primary fields, we introduce an auxiliary $(N+1)$-point function $Z(y)$ where we insert the energy-momentum tensor,
\begin{align}
 Z = \left< \prod_{i=1}^N V_{\Delta_i}(z_i) \right> \quad , \quad Z(y) = \left< T(y) \prod_{i=1}^N V_{\Delta_i}(z_i) \right> \ .
\end{align}
$Z(y)$ is a meromorphic function of $y$, with poles at $y=z_i$, whose residues are given by eq. \eqref{eq:tvd} (using the commutativity of fields).
Moreover $T(y)$, and therefore also $Z(y)$, vanish in the limit $y\to \infty$. So $Z(y)$ is completely determined by its poles and residues,
\begin{align}
 Z(y) = \sum_{i=1}^N \left(\frac{\Delta_i}{(y-z_i)^2} +\frac{1}{y-z_i}\frac{\partial}{\partial z_i}\right) Z\ .
 \label{eq:zy}
\end{align}
But $T(y)$ does not just vanish for $y\to \infty$, it behaves as $O(\frac{1}{y^4})$.
So the coefficients of $y^{-1}, y^{-2}, y^{-3}$ in the large $y$ expansion of $Z(y)$ must vanish, 
\begin{align}
 \sum_{i=1}^N \partial_{z_i} Z = \sum_{i=1}^N \left(z_i \partial_{z_i} + \Delta_i\right) Z = \sum_{i=1}^N \left(z_i^2 \partial_{z_i} + 2\Delta_iz_i\right) Z = 0\ .
 \label{eq:gward}
\end{align}
These three equations are called global Ward identities. 
The global Ward identities determine how $Z$ behaves under global conformal transformations of the Riemann sphere,
\begin{align}
 \left< \prod_{i=1}^N  V_{\Delta_i}\left(\frac{az_i+b}{cz_i+d}\right) \right>
 = \prod_{i=1}^N (cz_i +d)^{2\Delta_i} \left< \prod_{i=1}^N V_{\Delta_i}(z_i) \right>\ .
 \label{eq:zgc}
\end{align}
Let us solve the global Ward identities in the cases of one, two, three and four-point functions. For a one-point function, we have 
\begin{align}
\partial_z \Big< V_\Delta(z)\Big> =  \Delta \Big< V_\Delta(z)\Big> = 0\quad \text{so that} \quad \Big< V_\Delta(z)\Big> \neq 0 \implies V_\Delta \propto V_{\langle 1,1\rangle}\ .
\end{align}
Similarly, in the case of two-point functions, we find 
\begin{align}
 \Big< V_{\Delta_1}(z_1)V_{\Delta_2}(z_2) \Big> \propto \delta_{\Delta_1,\Delta_2} (z_1-z_2)^{-2\Delta_1} \ .
 \label{eq:2pt}
\end{align}
So a two-point function can be non-vanishing only if the two fields have the same dimension.
For three-point functions, there are as many equations \eqref{eq:gward} as unknowns $z_1,z_2,z_3$, and therefore a unique solution with no constraints on $\Delta_i$,
\begin{align}
 \left< \prod_{i=1}^3 V_{\Delta_i}(z_i) \right> \propto (z_1-z_2)^{\Delta_3-\Delta_1-\Delta_2} (z_1-z_3)^{\Delta_2-\Delta_1-\Delta_3} (z_2-z_3)^{\Delta_1-\Delta_2-\Delta_3}\ ,
 \label{eq:3pt}
\end{align}
with an unknown proportionality coefficient that does not depend on $z_i$.
For four-point functions, the general solution can be written as 
\begin{align}
 \left< \prod_{i=1}^4 V_{\Delta_i}(z_i) \right> 
 = z_{13}^{-2\Delta_1} z_{23}^{\Delta_1-\Delta_2-\Delta_3+\Delta_4} z_{24}^{-\Delta_1-\Delta_2+\Delta_3-\Delta_4} z_{34}^{\Delta_1+\Delta_2-\Delta_3-\Delta_4} G\left(\frac{z_{12}z_{34}}{z_{13}z_{24}}\right)\ ,
 \label{eq:4pt}
\end{align}
where $z_{ij} = z_i - z_j$ and $G(z)$ is an arbitrary function of the cross-ratio $z$. 
So the three global Ward identities effectively reduce the four-point function to a function of one variable $G$ -- equivalently, we can set $z_2,z_3,z_4$ to fixed values, and recover the four-point function from its dependence on $z_1$ alone. 

\begin{exo}[Global conformal symmetry]
~\label{exo:4pt}
Solve the global Ward identities for two-, three- and four-point functions, and recover eqs. \eqref{eq:2pt}, \eqref{eq:3pt} and \eqref{eq:4pt} respectively. 
Defining $V_\Delta(\infty) = \lim_{z\to\infty} z^{2\Delta}V_\Delta(z) $, show that this is finite when inserted into a two- or three-point function. More generally, show that this is finite using the behaviour \eqref{eq:zgc} of correlation functions under $z\to -\frac{1}{z}$. 
 Show that
 \begin{align}
  G(z) = \Big< V_{\Delta_1}(z) V_{\Delta_2}(0)V_{\Delta_3}(\infty)V_{\Delta_4}(1) \Big>\ .
 \end{align}
\end{exo}

We have been studying global conformal invariance of correlation functions of primary fields, rather than more general fields. This was not only for making things simpler, but also because correlation functions of descendents can be deduced from correlation functions of primaries. For example,
\begin{align}
 \Big< L_{-2}V_{\Delta_1}(z_1) V_{\Delta_2}(z_2)\cdots \Big>
  &= \frac{1}{2\pi i}\oint_{z_1} \frac{dy}{y-z_1} Z(y)
  = -\frac{1}{2\pi i} \sum_{i=2}^N \oint_{z_i} \frac{dy}{y-z_1} Z(y)\ ,
  \\
  &  =\sum_{i=2}^N\left(\frac{1}{z_1-z_i}\frac{\partial}{\partial z_i} +\frac{\Delta_i}{(z_i-z_1)^2}\right) Z\ ,
  \label{eq:ltv}
\end{align}
where we used first eq. \eqref{eq:lvtv} for $L_{-2}V_{\Delta_1}(z_1)$, and then eq. \eqref{eq:zy} for $Z(y)$.
This can be generalized to any correlation function of descendent fields. The resulting equations are called local Ward identities.

\subsection{Belavin--Polyakov--Zamolodchikov equations}

Local and global Ward identities are all we can deduce from conformal symmetry. But correlation functions that involve degenerate fields obey additional equations. 

For example, let us replace $V_{\Delta_1}(z_1)$ with the degenerate primary field $V_{\langle 1, 1 \rangle}(z_1)$
in our $N$-point function $Z$. Since $\frac{\partial}{\partial z_1} V_{\langle 1, 1 \rangle}(z_1) = L_{-1} V_{\langle 1, 1 \rangle}(z_1) =0$,
we obtain $\frac{\partial}{\partial z_1} Z =0$. 
In the case $N=3$, having $\Delta_1=\Delta_{\langle 1,1\rangle}=0$ in the three-point function \eqref{eq:3pt} leads to
\begin{align}
 \left< V_{\langle 1, 1 \rangle}(z_1) V_{\Delta_2}(z_2) V_{\Delta_3}(z_3) \right> \propto (z_1-z_2)^{\Delta_3-\Delta_2} (z_1-z_3)^{\Delta_2-\Delta_3} (z_2-z_3)^{-\Delta_2-\Delta_3}\ , 
\end{align}
and further imposing $z_1$-independence leads to 
\begin{align}
 \left< V_{\langle 1, 1 \rangle}(z_1) V_{\Delta_2}(z_2) V_{\Delta_3}(z_3) \right> \neq 0 \quad \implies \quad \Delta_2=\Delta_3\ .
 \label{eq:vvvnz}
\end{align}
This coincides with the condition \eqref{eq:2pt} that the two-point function $\left<V_{\Delta_2}(z_2)V_{\Delta_3}(z_3)\right>$ does not vanish. Actually, the field $V_{\langle 1,1\rangle}$ is an identity field, i.e. a field whose presence does not affect correlation functions. (See Exercise \ref{exo:id}.)

In the case of $V_{\langle 2, 1 \rangle}(z_1)$, we have  
\begin{align}
\left(L_{-1}^2 + b^2 L_{-2}\right) V_{\langle 2, 1 \rangle}(z_1)  = 0\qquad \text{so that} \qquad L_{-2}V_{\langle 2, 1 \rangle}(z_1) = -\frac{1}{b^2}\frac{\partial^2}{\partial z_1^2} V_{\langle 2, 1 \rangle}(z_1)\ .
\end{align}
Using the local Ward identity \eqref{eq:ltv},
this leads to the second-order Belavin--Polyakov--Zamolodchikov partial differential equation
\begin{align}
 \left( \frac{1}{b^2}\frac{\partial^2}{\partial z_1^2} + \sum_{i=2}^N\left(\frac{1}{z_1-z_i}\frac{\partial}{\partial z_i} +\frac{\Delta_i}{(z_1-z_i)^2}\right) \right)\left< V_{\langle 2, 1 \rangle}(z_1) \prod_{i=2}^N V_{\Delta_i}(z_i) \right>  = 0\ .
 \label{eq:bpz}
\end{align}
More generally, a correlation function with the degenerate field $V_{\langle r,s\rangle}$ obeys a partial differential equation of order $rs$. 
\begin{exo}[Second-order BPZ equation for a three-point function]
 ~\label{exo:bpz3pt}
 Show that 
\begin{align}
 \left< V_{\langle 2, 1 \rangle} V_{\Delta_2} V_{\Delta_3} \right> \neq 0 \quad \implies \quad 
 P_2 = P_3 \pm \frac{b}{2}\ .
 \label{eq:alpm}
\end{align}
\end{exo}
In the case of a four-point function, the BPZ equation amounts to a differential equation for the function of one variable $G(z)$.

\begin{exo}[BPZ second-order differential equation]
 ~\label{exo:bpz}
 Show that the second-order BPZ equation for $G(z)=\Big< V_{\langle 2, 1 \rangle}(z) V_{\Delta_1}(0)V_{\Delta_2}(\infty)V_{\Delta_3}(1) \Big>$ is
 \begin{align}
  \left\{ \frac{z(1-z)}{b^2}\frac{\partial^2}{\partial z^2} + (2z-1){\frac{\partial}{\partial z}} +\Delta_{\langle 2,1 \rangle} +\frac{\Delta_1}{z}-\Delta_2 + \frac{\Delta_3}{1-z}\right\} G(z)=0\ ,
\label{eq:ode}
 \end{align}
\end{exo}

\section{Conformal bootstrap}

We have seen how conformal symmetry leads to linear equations for correlation functions: Ward identities and BPZ equations. 
In order to fully determine correlation functions, we need additional, nonlinear equations, and therefore additional axioms: single-valuedness of correlation functions, and existence of operator product expansions. 
Using these axioms for studying conformal field theories is called the conformal bootstrap method. 

\subsection{Single-valuedness}\label{sec:sv}

\begin{hyp}[Single-valuedness]
 ~\label{hyp:sv}
 Correlation functions are single-valued functions of the positions, i.e. they have trivial monodromies.
\end{hyp}
Our two-point function \eqref{eq:2pt} however has nontrivial monodromy unless $\Delta_1\in \frac12\mathbb{Z}$, as a  result of solving holomorphic Ward identities. 
We would rather have a single-valued function of the type $|z_1-z_2|^{-4\Delta_1} = (z_1-z_2)^{-2\Delta_1} (\bar z_1-\bar z_2)^{-2\Delta_1}$.
This suggests that we need antiholomorphic Ward identities as well, and therefore a second copy of the Virasoro algebra.

\begin{hyp}[Left and right Virasoro algebras]
 ~\label{hyp:lr}
 We have two mutually commuting Virasoro symmetry algebras with the same central charge, called left-moving or holomorphic, and right-moving or antiholomorphic. Their generators are written $L_n,\bar L_n$, with in particular
 \begin{align}
  \frac{\partial}{\partial z} V(z) = L_{-1}V(z)   \quad , \quad \frac{\partial}{\partial \bar z} V(z)= \bar L_{-1} V(z)   \ .
 \end{align}
 The generators of conformal transformations are the diagonal combinations $L_n+\bar L_n$.
\end{hyp}
Let us consider left- and right-primary fields $V_{\Delta_i,\bar\Delta_i}(z_i)$, with the 
two-point functions
\begin{align}
 \left<\prod_{i=1}^2 V_{\Delta_i,\bar\Delta_i}(z_i) \right> \propto \delta_{\Delta_1,\Delta_2}\delta_{\bar\Delta_1,\bar\Delta_2} (z_1-z_2)^{-2\Delta_1} (\bar z_1-\bar z_2)^{-2\bar\Delta_1}\ .
\end{align}
This is single-valued if and only if our two fields have half-integer spins,
\begin{align}
 \Delta -\bar \Delta \in \frac12\mathbb{Z}\ .
\end{align} 
The simplest case is $\Delta=\bar\Delta$, which leads to the definition

\begin{defn}[Diagonal states, diagonal fields and diagonal spectrums]
 ~\label{def:diag}
 A primary state or field is called diagonal if it has the same left and right conformal dimensions. A spectrum is called diagonal if all primary states are diagonal.
\end{defn}
For diagonal primary fields, we will now write  $V_\Delta(z) = V_{\Delta,\Delta}(z)$.

\subsection{Operator product expansion and crossing symmetry}

\begin{hyp}[Operator product expansion]
 ~\label{hyp:ope}
 Let $(|w_i\rangle)$ be a basis of the spectrum.
 There exist coefficients $C^i_{12}(z_1,z_2)$ such that we have the operator product expansion (OPE) 
 \begin{align}
  V_{|w_1\rangle}(z_1)V_{|w_2\rangle}(z_2) \underset{z_1\to z_2}{=} \sum_i C^i_{12}(z_1,z_2) V_{|w_i\rangle}(z_2)\ .
 \end{align}
 In a correlation function,
 this sum converges for $z_1$ sufficiently close to $z_2$.
\end{hyp}
OPEs allow us to reduce $N$-point functions to $(N-1)$-point functions, at the price of introducing OPE coefficients. 
Iterating, we can reduce any correlation function to a combination of OPE coefficients, and two-point functions. (We stop at two-point functions because they are simple enough for being considered as known quantities.) 

If
the spectrum is made of diagonal primary states and their descendent states, the OPE of two primary fields is
\begin{align}
 V_{\Delta_1}(z_1) V_{\Delta_2}(z_2) 
\underset{z_1\to z_2}{=} \sum_{\Delta\in S} C_{\Delta_1,\Delta_2,\Delta} |z_1-z_2|^{2(\Delta-\Delta_1-\Delta_2)}
 \Big(V_{\Delta}(z_2) + O(z_1-z_2) \Big)\ ,
 \label{eq:ope}
\end{align}
where the subleading terms are contributions of descendents fields. 
In particular, the $z_1,z_2$-dependence of the coefficients is dictated by the behaviour
of correlation functions under translations $z_i\to z_i+c$ and dilations $z_i\to\lambda z_i$, leaving a $z_i$-independent unknown factor $C_{\Delta_1,\Delta_2,\Delta}$.
Then, as in correlation functions, contributions of descendents are deduced from contributions of primaries via local Ward identitites.

\begin{exo}[Computing the OPE of primary fields]
~\label{exo:ope}
 Compute the first subleading term in the OPE \eqref{eq:ope}, and find
 \begin{align}
  O(z_1-z_2) = \frac{\Delta+\Delta_1-\Delta_2}{2\Delta} \Big( (z_1-z_2)L_{-1}+(\bar z_1-\bar z_2)\bar L_{-1}\Big) V_{\Delta}(z_2) + O((z_1-z_2)^2)\ .
 \end{align}
Hints: Insert $\oint_C dz(z-z_2)^2 T(z)$ on both sides of the OPE, for a contour $C$ that encloses both $z_1$ and $z_2$. Compute the relevant contour integrals with the help of eq. \eqref{eq:tvd}.
\end{exo}
\begin{exo}[$V_{\langle 1,1\rangle}$ is an identity field]
~\label{exo:id}
Using $\frac{\partial}{\partial z_1} V_{\langle 1,1\rangle}(z_1)=0$, show that the OPE of $V_{\langle 1,1\rangle}$ with another primary field is of the form 
\begin{align}
 V_{\langle 1,1\rangle}(z_1)V_\Delta(z_2) = C_\Delta V_\Delta(z_2)\ ,
\end{align}
where the subleading terms vanish. Inserting this OPE in a correlation function, show that the constant $C_\Delta$ actually does not depend on $\Delta$. Deduce that, up to a factor $C=C_\Delta$, the field $V_{\langle 1,1\rangle}$ is an identity field.
\end{exo}
Using the OPE, we can reduce a three-point function to a combination of two-point functions, and we find 
\begin{align}
 \left<\prod_{i=1}^3 V_{\Delta_i}(z_i) \right> = C_{\Delta_1,\Delta_2,\Delta_3} |z_1-z_2|^{2(\Delta_3-\Delta_1-\Delta_2)} |z_1-z_3|^{2(\Delta_2-\Delta_1-\Delta_3)} |z_2-z_3|^{2(\Delta_1-\Delta_2-\Delta_3)}\ ,
\end{align}
assuming the two-point function is normalized as $\left< V_{\Delta}(z_1)V_{\Delta}(z_2) \right> = |z_1-z_2|^{-4\Delta}$.
It follows that $C_{\Delta_1,\Delta_2,\Delta_3}$ coincides with the undertermined constant prefactor of the three-point function. This factor is called the three-point structure constant.
Let us now insert the OPE in a four-point function of primary fields:
\begin{align}
 \Big<V_{\Delta_1}(z)V_{\Delta_2}(0)V_{\Delta_3}(\infty)V_{\Delta_4}(1)\Big>
 &\underset{z\to 0}{=} \sum_{\Delta\in S} C_{\Delta_1,\Delta_2,\Delta} |z|^{2(\Delta-\Delta_1-\Delta_2)}
 \nonumber
\\ & \qquad \qquad \times 
 \left(\Big< V_\Delta(0)V_{\Delta_3}(\infty)V_{\Delta_4}(1)\Big> + O(z)\right) \ ,
 \\
 &\underset{z\to 0}{=} \sum_{\Delta\in S} C_{\Delta_1,\Delta_2,\Delta} C_{\Delta,\Delta_3,\Delta_4}
|z|^{2(\Delta-\Delta_1-\Delta_2)} \Big(1 +O(z) \Big)\ .
\end{align}
The contributions of descendents factorize into those of left-moving descendents, generated by the operators $L_{n<0}$, and right-moving descendents, generated by $\bar L_{n<0}$. So the last factor has a holomorphic factorization such that 
\begin{align}
\Big<V_{\Delta_1}(z)V_{\Delta_2}(0)V_{\Delta_3}(\infty)V_{\Delta_4}(1)\Big>
 =\sum_{\Delta\in S} C_{\Delta_1,\Delta_2,\Delta} C_{\Delta,\Delta_3,\Delta_4}  \mathcal{F}^{(s)}_\Delta(z) \mathcal{F}^{(s)}_\Delta(\bar z)\ .
 \label{sdec}
\end{align}

\begin{defn}[Conformal block]
 ~\label{def:block}
 The four-point conformal block on the sphere,
 \begin{align}
  \mathcal{F}^{(s)}_\Delta(z) \underset{z\to 0}{=} z^{\Delta-\Delta_1-\Delta_2}\Big( 1 + O(z) \Big)\ ,
  \label{eq:gsd}
 \end{align}
is the normalized contribution of the Verma module $\mathcal V_\Delta$ to a four-point function, obtained by summing over left-moving descendents. It is a locally holomorphic function of $z$. Its dependence on $c,\Delta_1,\Delta_2,\Delta_3,\Delta_4$ are kept implicit. The label $(s)$ stands for $s$-channel.
\end{defn}
Conformal blocks are in principle known, as they are universal functions, entirely determined by conformal symmetry. 
This is analogous to characters of representations, also known as zero-point conformal blocks on the torus.

\begin{exo}[Computing conformal blocks]
 ~\label{exo:block}
 Compute the conformal block $ \mathcal{F}^{(s)}_\Delta(z)$ up to the order $O(z)$, and find
 \begin{align}
  \mathcal{F}^{(s)}_\Delta(z) \underset{z\to 0}{=} z^{\Delta-\Delta_1-\Delta_2}\left( 1 + \frac{(\Delta+\Delta_1-\Delta_2)(\Delta+\Delta_4-\Delta_3)}{2\Delta}z + O(z^2) \right)\ .
  \label{eq:fsl}
 \end{align}
 Show that the first-order term has a pole when the Verma module $\mathcal{V}_\Delta$ has a null vector at level one.
 Compute the residue of this pole. Compare the condition that this residue vanishes with the condition \eqref{eq:vvvnz} that three-point functions involving $V_{\langle 1,1\rangle}$ exist.
\end{exo}

Our axiom \ref{hyp:ass} on the commutativity of fields implies that the OPE is associative, and that we can use the OPE of any two fields in a four-point function. In particular, using the OPE of the first and fourth fields, we obtain 
\begin{align}
 \Big<V_{\Delta_1}(z)V_{\Delta_2}(0)V_{\Delta_3}(\infty)V_{\Delta_4}(1)\Big>
 =\sum_{\Delta\in S} C_{\Delta,\Delta_1,\Delta_4} C_{\Delta_2,\Delta_3,\Delta}   \mathcal{F}^{(t)}_\Delta(z) \mathcal{F}^{(t)}_\Delta(\bar z)\ ,
 \label{tdec}
\end{align}
where $\mathcal{F}^{(t)}_\Delta(z) \underset{z\to 1}{=} (z-1)^{\Delta-\Delta_1-\Delta_4}\Big(1+O(z-1)\Big)$ is a $t$-channel conformal block.
The equality of our two decompositions \eqref{sdec} and \eqref{tdec} of the four-point function is called crossing symmetry, schematically 
\begin{align}
 \sum_{\Delta_s\in S} C_{12s} C_{s34} 
 \begin{tikzpicture}[baseline=(current  bounding  box.center), very thick, scale = .3]
\draw (-1,2) node [left] {$2$} -- (0,0) -- node [above] {$s$} (4,0) -- (5,2) node [right] {$3$};
\draw (-1,-2) node [left] {$1$} -- (0,0);
\draw (4,0) -- (5,-2) node [right] {$4$};
\end{tikzpicture} 
= \sum_{\Delta_t\in S} C_{23t}C_{t41} 
\begin{tikzpicture}[baseline=(current  bounding  box.center), very thick, scale = .3]
 \draw (-2,3) node [left] {$2$} -- (0,2) -- node [left] {$t$} (0,-2) -- (-2, -3) node [left] {$1$};
\draw (0,2) -- (2,3) node [right] {$3$};
\draw (0,-2) -- (2, -3) node [right] {$4$};
\end{tikzpicture}
\ .
\label{csd}
\end{align}
The unknowns in this equation are the spectrum $S$ and three-point structure constant $C$. 
Any solution such that $C$ is invariant under permutations allows us to consistently compute arbitrary correlation functions on the sphere \cite{ms89b}, not just four-point functions.

\begin{defn}[Conformal field theory]
~\label{def:cft}
A (model of) conformal field theory on the Riemann sphere is a spectrum $S$ and a permutation-invariant three-point structure constant $C$ that obey crossing symmetry.
\end{defn}

\begin{defn}[Defining and solving]
 ~\label{def:def}
 To define a conformal field theory is to give principles that uniquely determine its spectrum $S$ and correlation functions $\left<\prod_{i=1}^N V_{|w_i\rangle}(z_i)\right>$ with $|w_i\rangle\in S$.
 To solve a conformal field theory is to actually compute them.
\end{defn}

\subsection{Degenerate fields and the fusion product}\label{sec:dffp}

Crossing symmetry equations are powerful, but typically involve infinite sums, which makes them difficult to solve.
However, if at least one field is degenerate, then the four-point function belongs to the finite-dimensional space of solutions of a BPZ equation, and is therefore a combination of finitely many conformal blocks. 
For example,
$G(z)=\Big< V_{\langle 2, 1 \rangle}(z) V_{\Delta_1}(0)V_{\Delta_2}(\infty)V_{\Delta_3}(1) \Big>$ is a combination of only two holomorphic $s$-channel conformal blocks.
These two blocks are a particular basis of solutions of the BPZ equation \eqref{eq:ode}.
They are fully characterized by their asymptotic behaviour near $z=0$ \eqref{eq:gsd}, where the BPZ equation allows only two values of $\Delta$, namely $\Delta\in\{\Delta(P_1-\frac{b}{2}),\Delta(P_1+\frac{b}{2})\}$.
Another basis of solutions of the same BPZ equation is given by two $t$-channel blocks, which are characterized by their power-like behaviour near $z=1$.
\begin{align}
 \mathcal{F}^{(s)}_\pm(z)  =  
 \begin{tikzpicture}[baseline=(current  bounding  box.center), very thick, scale = .35]
\draw (-1,2) node [left] {$P_1$} -- (0,0) -- node [above] {$P_1\pm \frac{b}{2}$} (4,0) -- (5,2) node [right] {$P_2$};
\draw[dashed] (-1,-2) node [left] {$\langle 2,1 \rangle$} -- (0,0);
\draw (4,0) -- (5,-2) node [right] {$P_3$};
\end{tikzpicture}
\qquad , \qquad 
\mathcal{F}^{(t)}_\pm(z)  =  
 \begin{tikzpicture}[baseline=(current  bounding  box.center), very thick, scale = .35]
 \draw (-2,3) node [left] {$P_1$} -- (0,2) -- node [left] {$P_3\pm \frac{b}{2}$} (0,-2);
 \draw[dashed] (0, -2) -- (-2, -3) node [left] {$\langle 2,1 \rangle$};
\draw (0,2) -- (2,3) node [right] {$P_2$};
\draw (0,-2) -- (2, -3) node [right] {$P_3$};
\end{tikzpicture}
\label{gpic}
\end{align}
These blocks are written in terms of the hypergeometric function,
\begin{multline}
\renewcommand{\arraystretch}{1.3}
 \mathcal{F}^{(s)}_\epsilon(z) = z^{b(\frac{Q}{2}-\epsilon P_1)} (1-z)^{b(\frac{Q}{2}+P_3)} 
 \\
 \times F\left(\tfrac12 + b(-\epsilon P_1+P_2+P_3), \tfrac12 + b(-\epsilon P_1-P_2+P_3), 1 - 2b\epsilon P_1, z\right)\ ,
\label{gpm}
\end{multline}
\begin{multline}
 \mathcal{F}^{(t)}_\eta(z) = z^{b(\frac{Q}{2}+P_1)} (1-z)^{b(\frac{Q}{2}-\eta P_3)}
 \\
 \times F\left(\tfrac12 + b(P_1+P_2-\eta P_3), \tfrac12 + b(P_1-P_2-\eta P_3), 1 - 2b\eta P_3, 1-z\right)\ .
 \label{tpm}
\end{multline}
Let us build single-valued four-point functions as linear combinations of such blocks. Single-valuedness near $z=0$ forbids terms such as $\mathcal{F}^{(s)}_{-}(z) \mathcal{F}^{(s)}_{+}(\bar z)$, and we must have 
\begin{align}
 G(z) = \sum_{\epsilon=\pm} c^{(s)}_{\epsilon} \mathcal{F}^{(s)}_{\epsilon}(z) \mathcal{F}^{(s)}_{\epsilon}(\bar z) = \sum_{\eta=\pm} c^{(t)}_{\eta} \mathcal{F}^{(t)}_{\eta}(z) \mathcal{F}^{(t)}_{\eta}(\bar z)\ .
 \label{gz}
\end{align}
The $s$- and $t$-channel blocks are two bases of the same space of solutions of the BPZ equation, and they are linearly related,
\begin{align}
 \mathcal{F}^{(s)}_{\epsilon}(z) = \sum_{\eta=\pm} F_{\epsilon\eta} \mathcal{F}^{(t)}_{\eta}(z)\ .
\end{align}
In particular, this implies 
\begin{align}
 \frac{c_{+}^{(s)}}{c_{-}^{(s)}} = -\frac{F_{-+}F_{--}}{F_{++}F_{+-}} \ .
 \label{eq:coc}
\end{align}
We will later express $c_\pm^{(s)}$ in terms of three-point structure constants, and obtain equations for these structure constants.

The presence of only two $s$-channel fields with momentums $P_1\pm \frac{b}{2}$ means that the operator product expansion $V_{\langle 2, 1 \rangle}(z) V_{P_1}(0)$ involves only two primary fields $V_{P_1\pm \frac{b}{2}}(0)$. 

\begin{defn}[Fusion product]
 ~\label{def:fus}
 The fusion product is a bilinear product of representations of the Virasoro algebra, that encodes the constraints on OPEs from Virasoro symmetry and null vectors. In particular,
 \begin{align}
  \mathcal{R}_{\langle 1,1\rangle}\times \mathcal V_P = \mathcal V_P \quad , \quad 
  \mathcal{R}_{\langle 2,1\rangle}\times \mathcal V_P = \sum_\pm \mathcal V_{P\pm \frac{b}{2}}\quad , \quad  
  \mathcal{R}_{\langle 1,2\rangle}\times \mathcal V_P = \sum_\pm \mathcal V_{P\pm \frac{1}{2b}}\ .
  \label{eq:rv}
 \end{align}
 From the commutativity of fields, it follows that the fusion product is commutative and associative.
\end{defn}
The fusion product can be defined algebraically \cite{gab99}: the fusion product of two representations is a coset of their tensor product, where however the Virasoro algebra does not act as it would in the tensor product. (In the tensor product, central charges and conformal dimensions add.) 

Using the associativity of the fusion product, we have 
\begin{align}
 \mathcal{R}_{\langle 2,1\rangle}\times \mathcal{R}_{\langle 2,1\rangle}  \times \mathcal V_P  =
\mathcal{R}_{\langle 2,1\rangle}\times  \left(\sum_\pm \mathcal V_{P\pm \frac{b}{2}}\right) =
\mathcal V_{P - b} + 2\cdot \mathcal V_P + \mathcal V_{P + b} \ .
\end{align}
Since the fusion product of $\mathcal{R}_{\langle 2,1\rangle}\times \mathcal{R}_{\langle 2,1\rangle} $ with $\mathcal V_P$ has finitely many terms, $\mathcal{R}_{\langle 2,1\rangle}\times \mathcal{R}_{\langle 2,1\rangle} $
must be a degenerate representation. 
On the other hand, eq. \eqref{eq:rv} implies that $\mathcal{R}_{\langle 2,1\rangle}\times \mathcal{R}_{\langle 2,1\rangle} $ is made of representations with momentums $P_{\langle 2,1\rangle} \pm \frac{b}{2} = P_{\langle 1,1\rangle}, P_{\langle 3,1\rangle}$. Therefore,
\begin{align}
 \mathcal{R}_{\langle 2,1\rangle}\times \mathcal{R}_{\langle 2,1\rangle} = \mathcal{R}_{\langle 1,1\rangle} + \mathcal{R}_{\langle 3,1\rangle} \quad , \quad \mathcal{R}_{\langle 3,1\rangle} \times \mathcal V_P = \mathcal V_{P - b} + \mathcal V_P + \mathcal V_{P + b}\ .
\end{align}
It can be checked that $\mathcal{R}_{\langle 3,1\rangle}$ has a vanishing null vector at level $3$, so that our definition of $\mathcal{R}_{\langle 3,1\rangle}$ from fusion agrees with the definition from representation theory in Section \ref{sec:nv}.

\begin{exo}[Higher degenerate representations]
~\label{exo:hdr}
 By recursion on $r,s\in \mathbb{N}^*$, show that there exist degenerate representations $\mathcal{R}_{\langle r,s \rangle}$ with momentums $P_{\langle r,s \rangle}$ \eqref{eq:ars},
such that 
 \begin{align}
 \mathcal{R}_{\langle r,s \rangle}\times \mathcal{V}_P &= \mathcal{R}_{\langle r,s \rangle}\times \mathcal{V}_P = \sum_{i=-\frac{r-1}{2}}^{\frac{r-1}{2}} \sum_{j=-\frac{s-1}{2}}^{\frac{s-1}{2}} \mathcal{V}_{P + ib+jb^{-1}}\ ,
\label{rtv}
 \\
 \mathcal{R}_{\langle r_1,s_1 \rangle} \times \mathcal{R}_{\langle r_2,s_2 \rangle} &= \sum_{r_3\overset{2}{=}|r_1-r_2|+1}^{r_1+r_2-1}\ \sum_{s_3\overset{2}{=}|s_1-s_2|+1}^{s_1+s_2-1} \mathcal{R}_{\langle r_3,s_3 \rangle}\ ,
\label{rrsr}
\end{align}
where the sums run by increments of $2$ if there is a superscript in $\overset{2}{=}$, and $1$ otherwise.
\end{exo}
These fusion products will play a crucial role in minimal models, whose spectrums are made of degenerate representations. 

\section{Minimal models}

\begin{defn}[Minimal model]
 ~\label{def:mm}
 A minimal model is a conformal field theory whose spectrum is made of finitely many irreducible representations of the product of the left and the right Virasoro algebras.
\end{defn}

\subsection{Diagonal minimal models}\label{sec:amm}

We first focus on diagonal minimal models, whose spectrums are of the type
\begin{align}
 S = \bigoplus_\mathcal{R} \mathcal{R}\otimes  \mathcal{\bar R}\ ,
\end{align}
where $\mathcal{R}$ and $ \mathcal{\bar R}$ denote the same Virasoro representation, viewed as a representation of the left- or right-moving Virasoro algebra respectively.

\begin{hyp}[Degenerate spectrum]
 ~\label{hyp:deg}
 All representations that appear in the spectrum of a minimal model are degenerate.
\end{hyp}
It is natural to use degenerate representations, because in an OPE of degenerate fields, only finitely many representations can appear. Conversely, we now assume that all representations that are allowed by fusion do appear in the spectrum, in other words

\begin{hyp}[Closure under fusion]
 ~\label{hyp:stab}
 The spectrum is closed under fusion. 
\end{hyp}

Let us assume that the spectrum contains a nontrivial degenerate representation such as $\mathcal{R}_{\langle 2,1\rangle}$. Fusing it with itself, we get $\mathcal{R}_{\langle 1, 1\rangle}$ and $\mathcal{R}_{\langle 3,1\rangle}$. Fusing multiple times, we get $(\mathcal{R}_{\langle r, 1\rangle})_{r\in\mathbb{N}^*}$ due to $\mathcal{R}_{\langle 2,1\rangle} \times \mathcal{R}_{\langle r,1\rangle} = \mathcal{R}_{\langle r-1,1\rangle}  + \mathcal{R}_{\langle r+1,1\rangle}$. If the spectrum moreover contains $\mathcal{R}_{\langle 1,2\rangle}$, then it must contain all degenerate representations. 

\begin{defn}[Generalized minimal model]
 ~\label{def:gmm}
 For any value of the central charge $c\in\mathbb{C}$, the generalized minimal model is the conformal field theory whose spectrum is
 \begin{align}
  S^\mathrm{GMM} = \bigoplus_{r=1}^\infty \bigoplus_{s=1}^\infty \mathcal{R}_{\langle r,s \rangle}\otimes  \mathcal{\bar R}_{\langle r,s \rangle} \ ,
 \end{align}
 assuming it exists and is unique.
\end{defn}

So, using only degenerate representations is not sufficient for building minimal models.
In order to have even fewer fields in fusion products, let us consider representations that are multiply degenerate. For example, if $\mathcal{R}_{\langle 2, 1\rangle} = \mathcal{R}_{\langle 1, 3\rangle}$ has two vanishing null vectors, then $\mathcal{R}_{\langle 2, 1\rangle} \times \mathcal{R}_{\langle 2, 1\rangle} = \mathcal{R}_{\langle 1,1\rangle}$ has only one term, as the term $\mathcal{R}_{\langle 3, 1\rangle}$ is not allowed by the fusion rules of $\mathcal{R}_{\langle 1, 3\rangle}$.

In order for a representation to have two null vectors, we however need a coincidence of 
the type $\Delta_{\langle r,s \rangle} = \Delta_{\langle r',s' \rangle}$. 
This is equivalent to $P_{\langle r,s \rangle} \in \{ P_{\langle r',s' \rangle}, -P_{\langle r',s' \rangle}\}$, and it follows that
$b^2$ is rational,
\begin{align} 
 b^2 = - \frac{q}{p} \qquad \text{with} \qquad \left\{\begin{array}{l} (p,q)\in \mathbb{N}^*\times \mathbb{Z}^* \\ p, q\text{ coprime} \end{array} \right. 
 \qquad \text{i.e.} \qquad c = 1-6\frac{(q-p)^2}{pq}\ .
 \label{eq:bcmin}
\end{align}
For any integers $r,s$, we then have the coincidence 
\begin{align}
 \Delta_{\langle r,s \rangle} = \Delta_{\langle p-r, q-s\rangle}\ .
\end{align}
In particular, let the Kac table be the set $(r,s)\in [1, p-1]\times [1,q-1]$, and let us build a diagonal spectrum from the corresponding representations:
\begin{align}
 S_{p, q} = \frac12 \bigoplus_{r=1}^{p-1} \bigoplus_{s=1}^{q-1} \mathcal{R}_{\langle r,s \rangle}\otimes \mathcal{\bar{R}}_{\langle r,s \rangle}\ ,
\end{align}
where  $\mathcal{R}_{\langle r,s \rangle}=\mathcal{R}_{\langle p-r,q-s \rangle}$ now denotes a degenerate representation with two independent null vectors, and the factor $\frac12$ is here to avoid counting the same representation twice.
This spectrum is not empty provided the coprime integers $p,q$ are both greater than $2$,
\begin{align}
 p,q \geq 2 \ ,
 \label{eq:pqmin}
\end{align}
which implies in particular $b,Q\in i\mathbb{R}$ and $c<1$.

\begin{exo}[Closure of minimal model spectrums under fusion]
 ~\label{exo:cmm}
 Show that $S_{p,q}$ is closed under fusion, and that the fusion products of the representations that appear in $S_{p,q}$ are
 \begin{align}
  \mathcal{R}_{\langle r_1,s_1 \rangle} \times \mathcal{R}_{\langle r_2,s_2 \rangle} = \sum_{r_3\overset{2}{=}|r_1-r_2|+1}^{\min(r_1+r_2,2p-r_1-r_2)-1}\ \sum_{s_3\overset{2}{=}|s_1-s_2|+1}^{\min(s_1+s_2,2q-s_1-s_2)-1} \mathcal{R}_{\langle r_3,s_3 \rangle}\ .
\label{rrmm}
\end{align}
 Are all finite, nontrivial sets of multiply degenerate representations that close under fusion subsets of some $S_{p,q}$? Do such sets exist only if $p,q\geq 2$?
\end{exo}

\begin{defn}[Diagonal minimal model]
 ~\label{def:dmm}
 For $p,q\geq 2$ coprime integers, the A-series $(p,q)$ minimal model is the conformal field theory whose spectrum is $S_{p, q}$, assuming it exists and is unique.
\end{defn}
For example, the minimal model with the central charge $c=\frac12$ has the spectrum $S_{4,3}$, 
\begin{align}
\renewcommand{\arraystretch}{1.3}
 \left\{\begin{array}{l} \Delta_{\langle 1,1\rangle}=\Delta_{\langle 3,2\rangle} = 0 \ , \\ \Delta_{\langle 1,2\rangle} =\Delta_{\langle 3,1\rangle} = \frac12 \ , \\ \Delta_{\langle 2,1\rangle} =\Delta_{\langle 2,2\rangle} = \frac{1}{16} \ .\end{array}\right. 
 \qquad \iff \quad \text{the Kac table} \quad 
 \begin{array}{c|ccc} 2 & \frac{1}{2} & \frac{1}{16} & 0 \\ 1 & 0 & \frac{1}{16} & \frac{1}{2} \\  \hline & 1 & 2 & 3 \end{array} 
\end{align}

\subsection{D-series minimal models}\label{sec:dmm}

Let us look for non-diagonal minimal models. We therefore relax the assumption that fields be diagonal, and allow them to have integer spins. (We could allow half-integer spins, leading to fermionic minimal models \cite{pet88}.) We still assume that the spectrum is made of doubly degenerate representations, and is closed under fusion.

Given a rational value of $b^2=-\frac{q}{p}$, let us look for pairs of doubly degenerate representations whose dimensions differ by integers, using the identity
\begin{align}
 \Delta_{\langle p-r,s\rangle} -\Delta_{\langle r,s\rangle}= \left(r-\frac{p}{2}\right)\left(s-\frac{q}{2}\right)\ .
\end{align}
Without loss of generality we assume that $q$ is odd. Then we need $r-\frac{p}{2}$ to be an even integer, therefore $p$ is even and $r\equiv\frac{p}{2}\bmod 2$. Under these assumptions, the representation $\mathcal{R}_{\langle r,s\rangle}\otimes \bar{\mathcal{R}}_{\langle p-r,s\rangle}$ has integer spin. We now look for a spectrum whose non-diagonal sector is made of all representations of this type, for $(r,s)$ in the Kac table.

Fusing two such representations produces degenerate representations with odd values of $r$. If $p\equiv 0\bmod 4$, such representations do not belong to our non-diagonal sector, and must therefore be diagonal. We therefore build a diagonal sector from all indices $(r,s)$ in the Kac table with $r$ odd, not only if $p\equiv 0\bmod 4$, but also for $p\equiv 2\bmod 4$. 

\begin{defn}[D-series minimal model]
 For $p,q\geq 2$ coprime integers with $p\in 6+2\mathbb{N}$, the D-series $(p,q)$ minimal model is the conformal field theory whose spectrum is 
\begin{align}
 S_{p,q}^\text{D-series} = \frac12 \bigoplus_{r\overset{2}{=}1}^{p-1} \bigoplus_{s=1}^{q-1} \mathcal{R}_{ \langle r,s \rangle} \otimes \bar{\mathcal{R}}_{\langle r,s \rangle}\oplus \frac12\bigoplus_{\substack{1\leq r\leq p-1 \\ r\equiv \frac{p}{2}\bmod 2}} \bigoplus_{s=1}^{q-1} \mathcal{R}_{\langle r,s \rangle} \otimes \bar{\mathcal{R}}_{\langle p-r,s\rangle}\ ,
 \label{eq:sds}
\end{align}
assuming it exists and is unique.
\end{defn}
(We need $p\geq 6$ for our would-be non-diagonal sector to actually contain representations with nonzero spins.)

\begin{exo}[Fusion rules of D-series minimal models]
~\label{exo:frd}
 If $p\equiv 0\bmod 4$, show that the D-series minimal model's fusion rules are completely determined by the fusion rules of the corresponding Virasoro representations. Write these fusion rules, and show that they conserve diagonality, in the sense that any correlation function with an odd number of non-diagonal fields vanishes. Assuming conservation of diagonality still holds if $p\equiv 2\bmod 4$, write the fusion rules of all D-series minimal models.
\end{exo}

\section{Liouville theory}

\subsection{Definition}

\begin{defn}[Liouville theory]
 ~\label{def:liou}
 For any value of the central charge $c\in\mathbb{C}$, Liouville theory is the conformal field theory whose spectrum is 
 \begin{align}
  S^\mathrm{Liouville} 
= \int_{i\mathbb{R}_+}  dP\ \mathcal V_P \otimes 
   \bar{\mathcal V}_P\ , 
   \label{eq:sl}
 \end{align}
and whose correlation functions are smooth functions of $b$ and $P$, assuming it exists and its unique.
\end{defn}
Let us give some justification for this definition. We are looking for a diagonal theory whose spectrum is a continuum of representations of the Virasoro algebra. For $c\in \mathbb{R}$ it is natural to assume $\Delta\in \mathbb{R}$. Let us write this condition in terms of the momentum $P$,
\begin{align}
 \Delta \in \mathbb{R} \iff P\in \mathbb{R} \cup i\mathbb{R}\ ,
  \qquad
   \begin{tikzpicture}[scale = .4, baseline=(current  bounding  box.center)]
   \draw[-latex] (-1.6, 0)  -- (2, 0) node[below right] {$0$} -- (4, 0)  -- (6.3, 0) node[below] {$P$};
  \draw[red, ultra thick] (-1.6, 0) -- (5.5, 0);
  \draw[red, ultra thick] (2, -3) -- (2, 3);
 \end{tikzpicture}
\end{align}
From Axiom \ref{hyp:rep}, 
we need $\Delta$ to be bounded from below, and the natural bound is 
\begin{align}
 \Delta_\text{min}=\Delta\left(P=0\right) = \frac{Q^2}{4}=\frac{c-1}{24}\ .
\end{align}
This leads to $P \in i\mathbb{R}$. Assuming that each allowed representation appears only once in the spectrum, we actually restrict the momentums to $P \in i\mathbb{R}_+$, due to the reflection symmetry \eqref{eq:refm}.
We then obtain our guess \eqref{eq:sl} for the spectrum, equivalently $S^\mathrm{Liouville} 
= \int_{\frac{c-1}{24}}^\infty d\Delta\ \mathcal V_{\Delta}\otimes \bar{\mathcal{V}}_\Delta $.
We take this guess to hold not only for $c\in\mathbb{R}$, but also for $c\in\mathbb{C}$ by analyticity.

Other guesses for the lower bound may seem equally plausible, in particular $\Delta_\text{min}=0$. In the spirit of the axiomatic method, the arbiter for such guesses is the consistency of the resulting theory. 
This will be tested in Section \ref{sec:cs}, and the spectrum $S^\mathrm{Liouville}$ will turn out to be correct.

Let us schematically write two- and three-point functions in Liouville theory, as well as OPEs:
\begin{align}
 \Big< V_{P_1}V_{P_2} \Big>  &=  B(P_1)\delta(P_1-P_2)\ ,
 \label{eq:vv}
 \\
 \Big< V_{P_1}V_{P_2}V_{P_3} \Big> & = C_{P_1,P_2,P_3} \ ,
 \label{eq:vvv}
 \\
 V_{P_1}V_{P_2} &= \int_{i\mathbb{R}_+} dP\, \frac{C_{P_1,P_2,P}}{B(P)} \Big( V_P + \cdots\Big)\ ,
 \label{eq:v1v2}
\end{align}
where the expression for the OPE coefficient $\frac{C_{P_1,P_2,P}}{B(P)}$ in terms of two- and three-point structure constants is obtained by inserting the OPE into a three-point function. 
It would be possible to set $B(P)=1$ by renormalizing the primary fields $V_{P}$, but this would prevent $C_{P_1,P_2,P_3}$ from being a meromorphic function of the momentums, as we will see in Section \ref{sec:sol}.

In order to have reasonably simple crossing symmetry equations, we need degenerate fields. 
But the spectrum of Liouville theory is made of Verma modules, and does not involve any degenerate representations.
In order to have degenerate fields, we need a special axiom:

\begin{hyp}[Degenerate fields in Liouville theory]
 ~\label{hyp:degl}
 The degenerate fields $V_{\langle r, s\rangle}$, and their correlation functions, exist. 
\end{hyp}
By the existence of degenerate fields, we also mean that such fields and their correlation functions obey suitable generalizations of our axioms. 
In particular, we generalize Axiom \ref{hyp:ope} by assuming that there exists an OPE between the degenerate field $V_{\langle 2, 1\rangle}$, and a field $V_P$. 
However, according to the fusion rules \eqref{eq:rv}, this OPE leads to fields with momentums $P\pm \frac{b}{2}$, and in general
$P \in i\mathbb{R} \centernot\implies (P\pm\frac{b}{2}) \in i\mathbb{R}$.
We resort to the assumption in Definition \ref{def:liou} that correlation functions are smooth functions of $P$, and take $V_P$ to actually be defined for $P\in\mathbb{C}$ by analytic continuation. This allows us to write the OPE
\begin{align}
 V_{\langle 2, 1\rangle} V_P \sim C_-(P) V_{P-\frac{b}{2}} + C_+(P)V_{P +\frac{b}{2}}\ ,
 \label{degope}
\end{align}
where we introduced the degenerate OPE coefficients $C_\pm(P)$.

\subsection{Three-point structure constants}\label{sec:sol}

Let us determine the three-point structure constant by solving crossing symmetry equations. We begin with the equations that come from four-point functions with degenerate fields. These equations are enough for uniquely determining the three-point structure constant.

Let us determine the coefficients $c^{(s)}_{\epsilon}$ in the expression \eqref{gz} for the four-point function $\left\langle V_{\langle 2,1 \rangle}(x)V_{P_1}(0)V_{P_2}(\infty)V_{P_3}(1)\right\rangle$.
Using the degenerate OPE \eqref{degope} and the three-point function \eqref{eq:vvv}, we find
\begin{align}
\begin{tikzpicture}[baseline=(current  bounding  box.center), very thick, scale = .6]
\draw (-1,2) node [left] {$P_1$} -- (0,0) -- node [above] {$P_1+\epsilon\frac{b}{2}$} (4,0) -- (5,2) node [right] {$P_2$};
\draw (-1,-2) node [left] {$\langle 2,1\rangle$} -- (0,0);
\draw (4,0) -- (5,-2) node [right] {$P_3$};
\node at (1.5, -4) {$c_{\epsilon}^{(s)} = \colorboxed{red}{C_\epsilon(P_1)}\, \colorboxed{red}{C_{P_1+\epsilon\frac{b}{2},P_2,P_3}}  $};
\draw[dashed, ->, red] (.7,-3.2) to [out=90, in=-70] (.2, -.3);
\draw[dashed, ->, red] (3.5,-3.2) to [out=90, in=-110] (3.8, -.3);
\end{tikzpicture} 
\label{cs}
\end{align}
Crossing symmetry and single-valuedness of the four-point function imply that the two structure constants $c_\pm^{(s)}$ obey eq. \eqref{eq:coc},
\begin{align}
 \frac{C_+(P_1) C_{P_1+\frac{b}{2},P_2,P_3}}{C_-(P_1) C_{P_1-\frac{b}{2},P_2,P_3} } 
 =\gamma(2bP_1) \gamma(1+2bP_1)\prod_{\pm,\pm} \gamma(\tfrac12 -bP_1 \pm bP_2 \pm bP_3)\ ,
 \label{eq:shift}
\end{align}
where we introduce the ratio of Euler Gamma functions
\begin{align}
 \gamma(x) = \frac{\Gamma(x)}{\Gamma(1-x)}\ .
\label{gx}
\end{align}
In order to find the three-point structure constant $C_{P_1,P_2,P_3}$, we need to constrain the degenerate OPE coefficients $C_\epsilon(P)$. To do this, we consider the special case where the last field is degenerate too, i.e. the four-point function $\Big\langle V_{\langle 2,1 \rangle}(z) V_P(0) V_{P}(\infty) V_{\langle 2,1 \rangle}(1)\Big\rangle$.
In this case, using the degenerate OPE \eqref{degope} twice, and the two-point function \eqref{eq:vv}, we find
\begin{align}
\begin{tikzpicture}[baseline=(current  bounding  box.center), very thick, scale = .6]
\draw (-1,2) node [left] {$P$} -- (0,0) -- node [above] {$P+\epsilon\frac{b}{2}$} (4,0) -- (5,2) node [right] {$P$};
\draw (-1,-2) node [left] {$\langle 2,1\rangle$} -- (0,0);
\draw (4,0) -- (5,-2) node [right] {$\langle 2,1\rangle$};
\node at (1, -4) {$c_{\epsilon}^{(s)} = \colorboxed{red}{C_\epsilon(P)}\, \colorboxed{red}{B(P+\epsilon\tfrac{b}{2})}\, \colorboxed{red}{C_\epsilon(P)} $};
\draw[dashed, ->, red] (-.7,-3.2) to [out=90, in=-70] (.2, -.3);
\draw[dashed, ->, red] (4.6,-3.2) to [out=90, in=-110] (3.8, -.3);
\draw[dashed, ->, red] (1.9,-3.2) to [out=90, in=-110] (2, -.3);
\end{tikzpicture} 
\end{align}
Then eq. \eqref{eq:coc} boils down to 
\begin{align}
 \frac{C_+(P)^2B(P+\tfrac{b}{2})}{C_-(P)^2 B(P-\tfrac{b}{2})}
 =  \frac{\gamma(2bP)}{\gamma(-2bP)}
 \frac{\gamma(-b^2-2bP)}{\gamma(-b^2+2bP)} \ .
 \label{eq:shiftd}
\end{align}
Moreover, if we had the degenerate field $V_{\langle 1,2\rangle}$ instead of $V_{\langle 2,1\rangle}$ in our four-point functions, we would obtain the equations \eqref{eq:shift} and \eqref{eq:shiftd} with $b\to \frac{1}{b}$. Next, we will solve these equations.

In order to solve the shift equations for $C_{P_1,P_2,P_3}$, we need a function that produces Gamma functions when its argument is shifted by $b$ or $\frac{1}{b}$. More precisely, we need a function such that 
\begin{align}
  \frac{\Upsilon_b(x+b)}{\Upsilon_b(x)} = b^{1-2bx} \gamma(bx)\qquad \text{and} \qquad \frac{\Upsilon_b(x+\frac{1}{b})}{\Upsilon_b(x)} = b^{\frac{2x}{b}-1} \gamma(\tfrac{x}{b})\ ,
\label{upup}
\end{align}
where the prefactors ensure that the two shift equations are compatible with one another. If it exists and is continuous, this function must be unique (up to a constant factor) if $b^2\in \mathbb{R}_+-\mathbb{Q}$, because the ratio of two solutions would be a continuous function with aligned periods $b$ and $\frac{1}{b}$. In the complex plane, the periods $b$ and $\frac{1}{b}$ indeed look as follows:
\begin{equation}
 \begin{tikzpicture}[baseline=(current  bounding  box.center), scale = .6]
\draw (0, 2) node[left]{$i$} -- (0, 1) node[below left] {$0$} -- (1, 1) node[below] {$1$};
\draw [red, thick, latex-latex] (11,3) -- (11,1) node[fill, circle, minimum size = 1mm, inner sep = 0]{} -- (11,-.3);
\draw [red, thick, latex-latex] (8,3) -- (7,1) node[fill, circle, minimum size = 1mm, inner sep = 0]{}-- (7.6,-.2);
\draw [red, thick, latex-latex] (5,1) -- (3,1) node[fill, circle, minimum size = 1mm, inner sep = 0]{} -- (4.3,1) ;
\node at (4, -1.5){$\begin{array}{c} b^2>0 \\ c\geq 25 \end{array}$};
\node at (7.5, -1.5){$\begin{array}{c} b\in \mathbb{C} \\ c\in\mathbb{C} \end{array}$};
\node at (11, -1.5){$\begin{array}{c} b^2<0 \\ c\leq 1 \end{array}$};
 \end{tikzpicture}
\end{equation}
For $b^2\in \mathbb{R}_--\mathbb{Q}$, there is also a unique solution of the shifts equations that are obtained from eq. \eqref{upup} by $b^{\cdots} \to (ib)^{\cdots}$, namely 
\begin{align}
 \hat{\Upsilon}_b(x) = \frac{1}{\Upsilon_{ib}(-ix+ib)}\ .
\end{align}

\begin{exo}[Upsilon function]
~\label{exo:upsilon}
 For $b>0$, show that the solution of the shift equations \eqref{upup} is 
 \begin{align}
 \Upsilon_b(x) = \lambda_b^{(\frac{Q}{2}-x)^2}\prod_{m,n=0}^\infty f\left(\frac{\frac{Q}{2}-x}{\frac{Q}{2}+mb+nb^{-1}}\right) \quad \text{with} \quad f(x)=(1-x^2)e^{x^2}\ ,
 \label{eq:up}
\end{align}
 where $\lambda_b$ is a function of $b$ to be determined. Deduce that $\Upsilon_b(x)$ is holomorphic and obeys $\Upsilon_b(x)=\Upsilon_b(Q-x)$. By analyticity in $b$, deduce that $\Upsilon_b(x)$ and $\hat\Upsilon_b(x)$ can be defined for $\Re b^2>0$ and $\Re b^2<0$ respectively.
\end{exo}
Let us now solve the shift equation \eqref{eq:shift} using the function $\Upsilon_b$. We write the ansatz
\begin{align}
 C_{P_1,P_2,P_3} =  \frac{N_0 \prod_{i=1}^3 N(P_i)}{\prod_{\pm,\pm} \Upsilon_b\left(\tfrac{Q}{2}+P_1\pm P_2 \pm P_3\right)} \ ,
 \label{cppp}
\end{align}
where $N_0$ is a function of $b$, and $N(P)$ is a function of $b$ and $P$. 
The denominator of this ansatz takes care of the last factor of the shift equation, and we are left with an equation that involves the dependence on $P_1$ only, 
\begin{align}
 \frac{C_+(P_1) N(P_1+\frac{b}{2})}{C_-(P_1) N(P_1-\frac{b}{2}) } 
 =b^{-8bP_1}\gamma(2bP_1) \gamma(1+2bP_1) \ .
\end{align}
Combining this equation with the shift equation for $B(P)$ \eqref{eq:shiftd}, we can eliminate the unknown degenerate OPE coefficients $C_\pm(P)$, and we obtain
\begin{align}
 \frac{\left(N^2B^{-1}\right)(P+\frac{b}{2})}{\left(N^2B^{-1}\right)(P-\frac{b}{2})} = b^{-16bP} \frac{\gamma(2bP)}{\gamma(-2bP)} \frac{\gamma(-b^2+2bP)}{\gamma(-b^2-2bP)}\ .
 \label{nbs}
\end{align}
Together with its image under $b\to b^{-1}$, this equation has the solution
\begin{align}
 \left(N^2B^{-1}\right)(P) = \prod_\pm \Upsilon_b(\pm 2P)\ .
 \label{eq:ntbm}
\end{align}
Therefore, we have only determined the combination $N^2B^{-1}$, and not the individual functions $B$ and $N$ that appear in the two- and three-point functions. This is because we still have the freedom of performing changes of field normalization $V_P(z) \to \lambda(P)V_P(z)$. Under such changes, we have $B\to \lambda^2B$ and $N\to \lambda N$, while
the combination $N^2B^{-1}$ is invariant. Invariant quantities are the only ones that we can determine without choosing a normalization, and the only ones that will be needed for checking crossing symmetry. 

It can nevertheless be convenient to choose a particular field normalization, if only to simplify notations:
\begin{itemize}
 \item $N(P)=1$ is the simplest choice.
 \item $N(P)=\Upsilon_b(2P)$ leads to the DOZZ formula for $C_{P_1,P_2,P_3}$ (after Dorn, Otto, A.
Zamolodchikov and Al. Zamolodchikov): it is the natural normalization in the functional integral approach to Liouville theory \cite{zz95}.
\item $B(P)=1$ causes $N(P)$ (and therefore $C_{P_1,P_2,P_3}$) to have square-root branch cuts. However, it is a natural normalization in the context of minimal models, where $b$ and $P_i$ take discrete values, and there is no notion of analyticity in these variables.
\end{itemize}

Our three-point structure constant \eqref{cppp} holds if
$c\notin ]-\infty, 1]$. 
On the other hand, doing the replacement $\Upsilon_b\to \hat\Upsilon_b$, we obtain a solution $\hat C$ that holds if  $c\notin [25,\infty[$, together with the corresponding functions $\hat B$ and $\hat N$.
The solution of the shift equations is unique if $b$ and $b^{-1}$ are aligned, i.e. if $b^2\in\mathbb{R}$. For generic values of the central charge, both $C$ and $\hat C$ are solutions, and there are actually infinitely many other solutions. In order to prove the existence and uniqueness of Liouville theory, we will have to determine which solutions lead to crossing-symmetric four-point functions.
In the case of (generalized) minimal models, the momentums $P_{\langle r,s\rangle}$ will belong to a lattice with periods $b$ and $\frac{1}{b}$, so the shift equations have a unique solution $C=\hat{C}$. (Actually $C$ has poles when $P_i$ take degenerate values, one should take the residues.)

\subsection{Crossing symmetry}\label{sec:cs}

We have found that Liouville theory is unique at least if $b^2\in\mathbb{R}$. We will now address the question of its existence. 

Using the $V_{P_1}V_{P_2}$ OPE \eqref{eq:v1v2}, let us write
the $s$-channel decomposition of a Liouville four-point function,
\begin{align}
\begin{tikzpicture}[baseline=(current  bounding  box.center), very thick, scale = .6]
\draw (-1,2) node [left] {$P_2$} -- (0,0) -- node [above] {$P$} (4,0) -- (5,2) node [right] {$P_3$};
\draw (-1,-2) node [left] {$P_1$} -- (0,0);
\draw (4,0) -- (5,-2) node [right] {$P_4$};
\node at (-2.5, -4) {$\Big< V_{P_1}(z) V_{P_2}(0) V_{P_3}(\infty) V_{P_4}(1)\Big> = {\displaystyle\int_{i\mathbb{R}_+}} dP\ \colorboxed{red}{\frac{C_{P_1,P_2,P}}{B(P)}}\, \colorboxed{red}{ C_{P,P_3, P_4}}\, \mathcal{F}_{P}^{(s)}(z) \mathcal{F}_{P}^{(s)}(\bar z)$};
\draw[dashed, ->, red] (.6,-2.8) to [out=90, in=-70] (.2, -.3);
\draw[dashed, ->, red] (3.4,-3.2) to [out=90, in=-110] (3.8, -.3);
\end{tikzpicture} 
\end{align}
(We have a similar expression with $B,C \to \hat B,\hat C$ whenever the solutions $\hat B, \hat C$ exist.)
Let us accept for a moment that Liouville theory is crossing-symmetric if $b^2\in\mathbb{R}$ i.e. $c\geq 25$ or $c\leq 1$. 
The integrand of our $s$-channel decomposition is well-defined, and analytic as a function of $b$, in the much larger regions $c\notin]-\infty,1]$ and $c\notin [25,\infty[$ respectively. 
If the integral itself was analytic as well, then crossing symmetry would hold in these regions by analyticity. 

In order to investigate the analytic properties of the integral, let us first extend the integration half-line to a line, $\int_{i\mathbb{R}_+} \to \frac12 \int_{i\mathbb{R}}$. This is possible because the integrand is invariant under $P \to -P$. 
Let us then study the singularities of the integrand.
We accept that the conformal blocks $\mathcal{F}_{P}^{(s)}(z)$ have poles when $P = P_{\langle r, s\rangle} $ \eqref{eq:ars}, the momentums for which the $s$-channel representation becomes reducible \cite{zz90}.
We now plot the positions of these poles (blue regions) relative to the integration line (red), depending on the central charge:
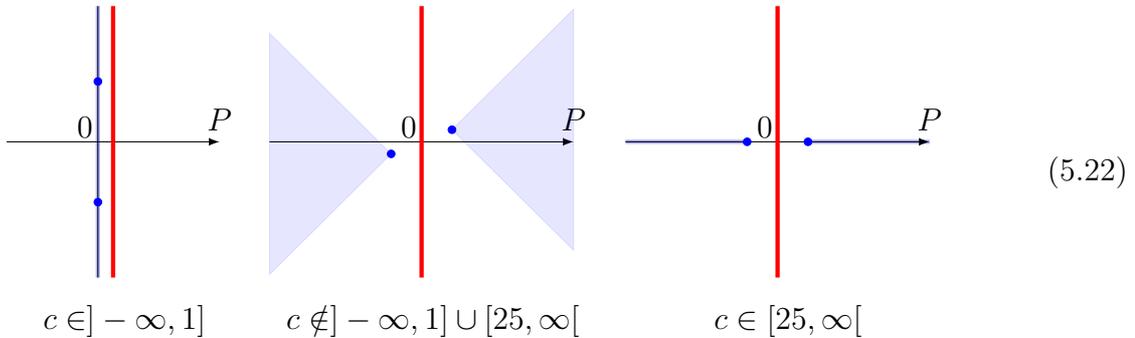
\begin{align}
 \newcommand{\polewedge}[3]{
\begin{scope}[#1]
\node[blue, draw,circle,inner sep=1pt,fill] at (0, 0) {};
\filldraw[opacity = .1, blue] (0,0) -- (4, -4) -- (4, 4) -- cycle;
\end{scope}
}
\begin{array}{ccc}
\begin{tikzpicture}[scale = .4, baseline=(current  bounding  box.center)]
  \draw[-latex] (-3,0) -- (0, 0) -- (4,0) node [above] {$P$};
  \draw[ultra thick, blue, opacity = .3] (0, -4.5) -- (0, 4.5);
  \draw (0, -4.5) -- (0, 4.5);
  \draw[ultra thick, red] (.5, -4.5) -- (.5, 4.5);
\node[above left] at (.2,-.2) {$0$};
\node[blue, draw,circle,inner sep=1pt,fill] at (0, 2) {};
\node[blue, draw,circle,inner sep=1pt,fill] at (0, -2) {};
 \end{tikzpicture}
 & 
 \begin{tikzpicture}[scale = .4, baseline=(current  bounding  box.center)]
  \draw[-latex] (-4,0)  --  (6,0) node [above] {$P$};
  \draw[ultra thick, red] (1, -4.5)  -- (1, 4.5);
  \node[above left] at (1.2,-.2) {$0$};
  \polewedge{rotate = 180, shift = {(0,.4)}}{$-\frac{Q}{2}$}{below};
  \polewedge{shift = {(2, .4)}}{$\frac{Q}{2}$}{above};
 \end{tikzpicture}
 &
 \begin{tikzpicture}[scale = .4, baseline=(current  bounding  box.center)]
 \draw[ultra thick, blue, opacity = .3] (0,0) -- (-4,0);
 \draw[ultra thick, blue, opacity = .3] (2,0) -- (6,0);
  \draw[-latex] (-4,0) -- (6,0) node [above] {$P$};
  \draw[ultra thick, red] (1, -4.5) -- (1, 4.5);
  \node[blue, draw,circle,inner sep=1pt,fill] at (0, 0) {};
\node[above left] at (1.2,-.2) {$0$};
\node[blue, draw,circle,inner sep=1pt,fill] at (2, 0) {};
 \end{tikzpicture}
 \vspace{3mm}
 \\
 c\in ]-\infty, 1] & c\notin ]-\infty, 1] \cup [25,\infty[ & c\in [25,\infty[
\end{array}
\end{align}
When $c$ varies in the region $c\notin ]-\infty,1]$, the poles never cross the integration line.
Therefore, 
the four-point function built from $C$ is analytic on $c\notin ]-\infty,1]$. 
So if Liouville theory exists for $c\geq 25$, then it also exists for $c\notin ]-\infty,1]$, with the same structure constant $C$. 
On the other hand, if $c\leq 1$, then the poles are on the integration line, and actually the line
has to be slightly shifted in order to avoid the poles. 
We cannot analytically continue the four-point function from the region $c\leq 1$ to complex values of $c$, because this would make infinitely many poles cross the integration line.
So the structure constant $\hat C$ is expected to be valid for $c\leq 1$ only.

That is how far we can easily get with analytic considerations. 
Let us now seek input from numerical tests of crossing symmetry, using Al. Zamolodchikov's recursive formula for computing conformal blocks \cite{zz90}. (See the associated \href{https://github.com/ribault/bootstrap-2d-Python/blob/master/Liouville_demo_2.ipynb}{Jupyter notebook}, and the article \cite{rs15}.)
We find that Liouville theory exists for all values of $c\in\mathbb{C}$, with the three-point structure constants $\hat C$ for $c\leq 1$, and $C$ otherwise.
We also find that generalized minimal models exist for all values of $c$, and minimal models exist at the discrete values \eqref{eq:bcmin} of $c\leq 1$ where they are defined. 
And we can numerically compute correlation functions with a good precision.

Historically, Liouville theory was first defined by quantizing a classical theory whose equation of motion is Liouville's equation. That definition actually gave its name to the theory. (That definition does not cover the case $c\leq 1$: using the name Liouville theory in this case, while natural in our approach, is not universally done at the time of this writing.) It can be shown that our definition of Liouville theory agrees with the historical definition, either by proving that the originally defined theory obeys our axioms, or by checking that both definitions lead to the same correlation functions, in particular the same three-point structure constants. See \cite{tes17} for a guide to the literature on the construction of Liouville theory by quantization. 

\appendix

\section{Side subjects}

\subsection{Free boson}\label{sec:fb}

\textit{This Appendix can be read after Section \ref{sec:cft}, and Exercise \ref{exo:fbs} can be done after Section \ref{sec:sv}.}

\vspace{2mm}

We will now introduce the conformal field theory of the free boson. This provides the simplest examples of a CFT with an extended symmetry algebra, i.e. an algebra that is strictly larger than the Virasoro algebra: the abelian affine Lie algebra. This is a good preparation for Wess--Zumino--Witten models, with their non-abelian affine Lie algebras.

Just as the Virasoro algebra can be defined by the self-OPE of the energy-momentum tensor \eqref{tt}, we define the abelian affine Lie algebra by the self-OPE of a locally holomorphic current $J(y)$,
\begin{align}
 J(y)J(z) \underset{y\to z}{=} \frac{-\frac12}{(y-z)^2} + O(1)\ .
\label{jj}
\end{align}
We are still doing conformal field theory, because we can build an energy-momentum tensor from the current. Actually, for any $Q\in\mathbb{C}$, we can build a Virasoro algebra, whose generators are the modes of 
\begin{align}
 T(y) = -(JJ)(y) - Q\partial J(y)\ .
\label{tqz}
\end{align}
We choose a value of $Q$, and assume that the resulting Virasoro algebra generates conformal transformations, and in particular obeys Axiom \ref{hyp:geom}. Given the behaviour \eqref{eq:tinf} of $T(y)$ at infinity, we assume 
\begin{align}
 J(y)\underset{y\to\infty}{=}\frac{Q}{y} +O\left(\frac{1}{y^2}\right) \ .
 \label{eq:ji}
\end{align}

\begin{defn}[Normal-ordered product]
 ~\label{def:nop}
 Given two locally holomorphic fields, their normal-ordered product is
 \begin{align}
 (AB)(z) = \frac{1}{2\pi i} \oint_z \frac{dy}{y-z} A(y)B(z)\ .
\label{abz}
\end{align}
Equivalently, if $\cunderbracket{A}{(y)}{B}(z)$ is the singular part of the OPE $A(y)B(z)$,
we have
\begin{align}
(AB)(z) &= \underset{y\to z}{\lim} \left(A(y)B(z)-\cunderbracket{A}{(y)}{B}(z)\right)\ ,
\\
 A(y)B(z) &= \cunderbracket{A}{(y)}{B}(z) + (AB)(z) + O(y-z)\ .
 \label{abope}
\end{align}
The normal-ordered product is neither associative, nor commutative.
\end{defn}
OPEs that involve normal-ordered products can be computed using Wick's theorem,
\begin{align}
 \cunderbracket{A}{(z)}{(BC)}(y) \underset{z\to y}{=} \frac{1}{2\pi i}\oint_y \frac{dx}{x-y}\left(\cunderbracket{A}{(z)}{B}(x)C(y) + B(x)\cunderbracket{A}{(z)}{C}(y)\right)\ .
\label{wick}
\end{align}
For example, we can compute 
\begin{align}
 (JJ)(y)J(z) \underset{z\to y}{=} -\frac{J(y)}{(y-z)^2} +O(1) \underset{y\to z}{=}  -{\frac{\partial}{\partial z}}\frac{J(z)}{y-z} + O(1)\ ,
\end{align}
which leads to 
\begin{align}
 T(y)J(z) \underset{y\to z}{=} \frac{-Q}{(y-z)^3} +{\frac{\partial}{\partial z}}\frac{1}{y-z}J(z) + O(1)\ .
\label{tqj}
\end{align}
Then we can compute the OPE $T(y)T(z)$, and we find the OPE \eqref{tt}, where the central charge is given by $c = 1+6 Q^2$ (repeating eq. \eqref{eq:cqb}). 

We define an affine primary field with the momentum $\alpha$ by the OPE 
\begin{align}
 J(y) V_\alpha(z) \underset{y\to z}{=} \frac{\alpha}{y-z} V_\alpha(z) + O(1)\ .
\label{jva}
\end{align}
Using Wick's theorem, we deduce 
\begin{align}
 T(y) V_\alpha(z) \underset{y\to z}{=} \frac{\alpha(Q-\alpha) }{(y-z)^2}V_\alpha(z) - \frac{2\alpha}{y-z} (JV_\alpha)(z) + O(1)\ . 
\end{align}
This means that our affine primary field is also a Virasoro primary field with the conformal dimension $\alpha(Q-\alpha)$, and that 
\begin{align}
 \partial V_\alpha(z) = -2\alpha (JV_\alpha)(z)\ .
 \label{eq:pv}
\end{align}
Knowing its poles and residues, we compute 
\begin{align}
 \left< J(y) \prod_{i=1}^N V_{\alpha_i}(z_i)\right> = \sum_{i=1}^N\frac{\alpha_i}{y-z_i} \left<\prod_{i=1}^N V_{\alpha_i}(z_i)\right>\ .
\end{align}
From eq. \eqref{eq:ji} we first deduce the global Ward identity 
\begin{align}
 \left(\textstyle\sum_{i=1}^N \alpha_i - Q\right) \left\langle \prod_{i=1}^N V_{\alpha_i}(z_i) \right\rangle = 0 \ ,
\end{align}
which means that the momentum is conserved. Then, using eq. \eqref{eq:pv}, we deduce 
\begin{align}
\left( {\frac{\partial}{\partial z_i}} +\sum_{j\neq i} \frac{2\alpha_i\alpha_j}{z_i-z_j} \right) \left\langle \prod_{i=1}^N V_{\alpha_i}(z_i) \right\rangle = 0 \ .
\label{kzl}
\end{align}
(This is the abelian version of the Knizhnik--Zamolodchikov equation.) 
We can solve this differential equation, and we find
\begin{align}
 \left\langle \prod_{i=1}^N V_{\alpha_i}(z_i) \right\rangle \propto \delta\left(\textstyle\sum_{i=1}^N \alpha_i - Q\right) \prod_{i<j} (z_i-z_j)^{-2\alpha_i\alpha_j}\ .
\end{align}
Affine symmetry determines the $z_i$-dependence of all correlation functions, while Virasoro symmetry did it only for two- and three-point functions. 

\begin{exo}[Free bosonic spectrums]
~\label{exo:fbs}
In the case $c=1$, we want to build CFTs with the abelian affine Lie algebra symmetry, whose primary states include diagonal states and non-diagonal states with integer spins. We interpret momentum conservation as implying that the spectrum is closed under the addition of momentums. Let $(\frac{i}{2R},\frac{i}{2R})$ be the left and right momentums of a diagonal primary state, and $(\alpha,\bar\alpha)$ the momentums of another state. Show that 
\begin{align}
 \frac{i}{R}(\alpha-\bar\alpha)\in\mathbb{Z}\ .
\end{align}
Under mild assumptions, deduce that the spectrum is of the type 
 \begin{align}
 S_R = \bigoplus_{(n,w)\in {\mathbb{Z}^2}} \mathcal{U}_{\frac{i}{2}\left(\frac{n}{R} + Rw\right)} \otimes \bar{\mathcal{U}}_{\frac{i}{2}\left(\frac{n}{R} - Rw\right)} \ ,
\label{sr}
\end{align}
where $\mathcal{U}_\alpha$ is the representation of the abelian affine Lie algebra that corresponds to the primary field $V_\alpha$. The corresponding CFT is called the compactified free boson, with the compactification radius $R$. Do compactifield free bosons exist for $c\neq 1$?
\end{exo}

\subsection{Modular bootstrap}\label{sec:mb}

\textit{This Appendix can be read after Section \ref{sec:amm}.}

\vspace{2mm}

The torus zero-point function (or partition function) is a correlation function that only depends on the spectrum, and on characters of representations of the Virasoro algebra. Since there is no dependence on three-point structure constants, and since characters are much simpler than four-point blocks, the torus partition function is much simpler than four-point functions on the sphere. Nevertheless, the partition function obeys a nontrivial constraint called modular invariance.

\begin{defn}[Modular bootstrap]
The modular bootstrap consists in using the modular invariance of the torus partition function for deriving constraints on the spectrum.
\end{defn}
However, modular invariant partition functions do not always correspond to consistent CFTs: consistency of a CFT on all Riemann surfaces is equivalent to crossing symmetry of the sphere four-point function and modular invariance of the torus one-point function \cite{ms89b}. And some CFTs are consistent on the sphere only.

\begin{defn}[Torus partition function]
 For a CFT with the spectrum $S$, the partition function on the torus $\frac{\mathbb{C}}{\mathbb{Z}+\tau \mathbb{Z}}$ is 
 \begin{align}
  Z(\tau) = \operatorname{Tr}_S q^{L_0-\frac{c}{24}}\bar q^{\bar L_0-\frac{c}{24}}\quad \text{where} \quad q=e^{2\pi i \tau}\ .
 \end{align}
\end{defn}

\begin{hyp}[Modular invariance]
For $\left(\begin{smallmatrix} a & b \\ c & d\end{smallmatrix}\right)\in SL_2(\mathbb{Z})$, the torus partition function is invariant under the corresponding modular transformation,
\begin{align}
 Z(\tau) = Z\left(\frac{a\tau+b}{c\tau +d} \right)\ .
\end{align}
\end{hyp}
Let us give some justification for the definition of the torus partition function. The energy operator on the complex plane is $L_0+\bar{L}_0$: under the conformal map $z\mapsto \log z$ to the infinite cylinder, this becomes $L_0+\bar{L}_0-\frac{c}{12}$. To get the torus, we truncate the cylinder to a finite length $i(\tau-\bar{\tau})$, and identify points on the upper and lower boundaries after a rotation by $\tau+\bar{\tau}$. The partition function is the trace of the operator that performs this identification.
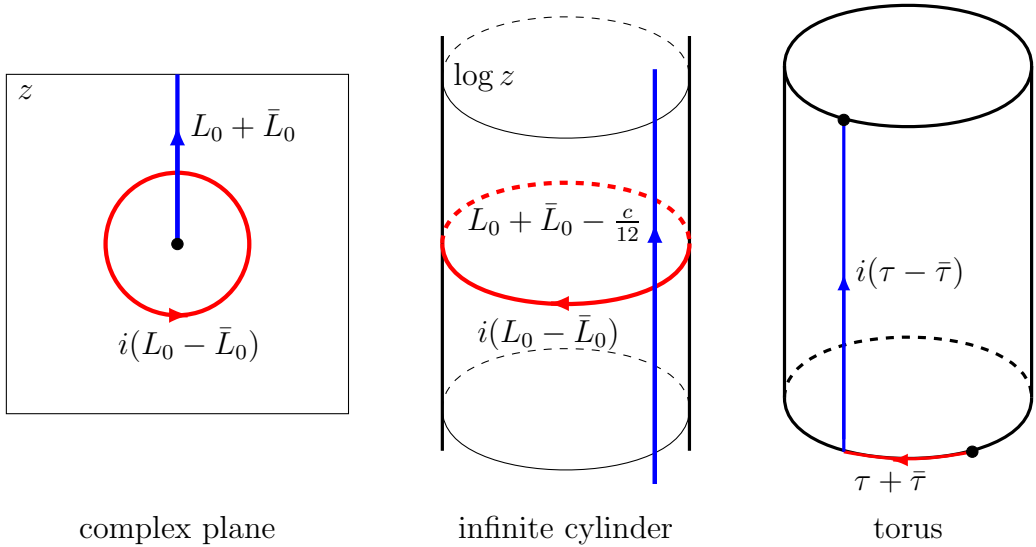
\begin{align*}
\renewcommand{\arraystretch}{1.3}
\renewcommand{\arraycolsep}{6mm}
\begin{array}{ccc}
 \begin{tikzpicture}[scale = .45, baseline=(current  bounding  box.center)]
  \draw (-5, -5) -- (5, -5) -- (5, 5) -- (-5, 5) node[below right]{$z$} -- cycle;
  \draw [red, ultra thick] (0, 0) circle (2.1);
  \draw [red, ultra thick, -latex] (.3, -2.1) -- (.35, -2.1) node[black, below] {$i(L_0-\bar{L}_0)$};
  \draw [blue, ultra thick] (0, 0) -- (0, 5);
  \draw [blue, ultra thick, -latex] (0, 0) -- (0, 3.5) node[black, right] {$L_0+\bar{L}_0$};
  \node[draw, circle, inner sep=1.5pt, fill] at (0, 0) {};
 \end{tikzpicture}
 & 
 \begin{tikzpicture}[xscale = .65, yscale = .55, baseline=(current  bounding  box.center)]
  \draw [very thick] (0, -5) -- (0, 5);
  \draw [very thick] (5, -5) -- (5, 5);
  \draw (0, -4) to [out = -90, in = -90] (5, -4);
  \draw [dashed] (0, -4) to [out = 90, in = 90] (5, -4);
  \draw (0, 4) node[right] {$\log z$} to [out = -90, in = -90] (5, 4);
  \draw [dashed] (0, 4) to [out = 90, in = 90] (5, 4);
  \draw [red, ultra thick] (0, 0) to [out = -90, in = -90] (5, 0);
  \draw [red, ultra thick, dashed] (0, 0) to [out = 90, in = 90] (5, 0);
  \draw [red, ultra thick, -latex] (2.2, -1.45) -- (2.15, -1.45) node[black, below] {$i(L_0-\bar{L}_0)$};
  \draw [blue, ultra thick] (4.3, -5.8) -- (4.3, 4.2);
  \draw [blue, ultra thick, -latex] (4.3, -5.8) -- (4.3, .5) node[black, left] {$L_0+\bar{L}_0-\frac{c}{12}$};
 \end{tikzpicture}
 & 
 \begin{tikzpicture}[xscale = .65, yscale = .55, baseline=(current  bounding  box.center)]
  \draw [very thick] (0, -4) -- (0, 4);
  \draw [very thick] (5, -4) -- (5, 4);
  \draw [very thick] (0, -4) to [out = -90, in = -90] (5, -4);
  \draw [very thick, dashed] (0, -4) to [out = 90, in = 90] (5, -4);
  \draw [very thick] (0, 4) to [out = -90, in = -90] (5, 4);
  \draw [very thick] (0, 4) to [out = 90, in = 90] (5, 4);
  \draw [red, very thick] (1.2, -5.3) to [out = -13, in = -167] (3.8, -5.3);
  \draw [red, very thick, -latex] (2.2, -5.5) -- (2.15, -5.5) node[black, below] {$\tau+\bar{\tau}$};
  \draw [blue, very thick] (1.2, -5.3) -- (1.2, 2.7);
  \draw [blue, very thick, -latex] (1.2, -5) -- (1.2, -1) node[black, right] {$i(\tau-\bar\tau)$};
  \node[draw, circle, inner sep=1.5pt, fill] at (3.8, -5.3) {};
  \node[draw, circle, inner sep=1.5pt, fill] at (1.2, 2.7) {};
 \end{tikzpicture}
 \\
 \text{complex plane}
 & 
 \text{infinite cylinder}
 &
 \text{torus}
 \end{array}
\end{align*}

Actually, modular invariance reduces to the invariance under two particular modular transformations,
\begin{align}
 T(\tau) = \tau + 1 \qquad , \qquad S(\tau) = -\frac{1}{\tau}\ .
\end{align}
The condition $Z(\tau)=Z(\tau+1)$ amounts to $L_0-\bar{L}_0$ having integer eigenvalues, in other words all states having integer conformal spins.
The condition $Z(\tau)=Z(-\frac{1}{\tau})$ is more complicated: to exploit this condition, let us decompose the spectrum into factorized representations of the type $\mathcal{R}\otimes \bar{\mathcal{R}}'$. 
The contribution of $\mathcal{R}\otimes \bar{\mathcal{R}}'$ to the partition function is $\chi_{\mathcal{R}}(\tau) \chi_{\mathcal{R}'}(-\bar\tau)$, where we define the character of a representation as
\begin{align}
 \chi_\mathcal{R}(\tau) = \operatorname{Tr}_\mathcal{R} q^{L_0-\frac{c}{24}}\ .
\end{align}
From $Z(\tau)=Z(-\frac{1}{\tau})$, we first deduce that there exists a modular $S$-matrix such that 
\begin{align}
 \chi_\mathcal{R}(\tau) = \sum_{\mathcal{R}'} S_{\mathcal{R},\mathcal{R}'} \chi_{\mathcal{R}'}(-\tfrac{1}{\tau})\ .
\end{align}
Let us consider a diagonal CFT, with the partition function $Z(\tau)=\sum_\mathcal{R} \chi_\mathcal{R}(\tau)\chi_{\mathcal{R}}(-\bar{\tau})$.
Comparing two expressions for $Z(-\frac{1}{\tau})$, 
\begin{align}
Z(-\tfrac{1}{\tau}) = \sum_{\mathcal{R}',\mathcal{R}''} \sum_{\mathcal{R}} S_{\mathcal{R},\mathcal{R}'} S_{\mathcal{R},\mathcal{R}''} \chi_{\mathcal{R}'}(-\tfrac{1}{\tau}) \chi_{\mathcal{R}''}(\tfrac{1}{\bar\tau}) = \sum_{\mathcal{R}} \chi_{\mathcal{R}}(-\tfrac{1}{\tau}) \chi_{\mathcal{R}}(\tfrac{1}{\bar\tau})\ ,
\end{align}
we deduce $SS^T=\text{Id}$. Since $S^2=\text{Id}$ by construction, this means that a diagonal modular invariant partition function exists if and only if the $S$-matrix is symmetric. (This reasoning must be modified in CFTs based on larger symmetry algebras \cite{fms97}. In particular, characters depend not just on $\tau$ but on extra variables, and $S^2$ is no longer identity but the charge conjugation matrix, where charge conjugation is the involution $\mathcal{R}\to \mathcal{R}^*$ such that $\left<V_{\mathcal{R}} V_{\mathcal{R}^*}\right>\neq 0$.)

Let us compute the characters and modular $S$-matrix of minimal models. We start with the character of a Verma module with momentum $P$,
\begin{align}
 \chi_P(\tau) = \frac{q^{-P^2}}{\eta(\tau)} \quad \text{with} \quad \eta(\tau) = q^{\frac{1}{24}} \prod_{n=1}^\infty (1-q^n)\ ,
\end{align}
where the nontrivial factor is called the Dedekind $\eta$ function. Why this function? If $L_{-1}$ was our only creation mode, we would have one state at each level, and the character would be $\chi(\tau)\sim 1+q+q^2+q^3+\cdots = \frac{1}{1-q}$. If we had $L_{-2}$ instead, the character would be $\chi(\tau)\sim \frac{1}{1-q^2}$. And in order to count the states that come from two creation modes, we must multiply the corresponding series.

In order to compute the character of a degenerate representation, we should subtract the contributions of null vectors and their descendent states. For a simply degenerate representation, the character is therefore
\begin{align}
 \chi_{\langle r,s\rangle}(\tau) = \chi_{P_{\langle r,s\rangle}}(\tau) - \chi_{P_{\langle r,-s\rangle}}(\tau)\ .
\end{align}
For a fully degenerate representation in the Kac table of the $(p, q)$ minimal model, the structure is a bit more complicated: we have to subtract the two null vectors and their descendents, but add again the intersection of their two Verma submodules, which was subtracted twice.
It turns out that 
the corresponding $S$-matrix is symmetric, showing that the A-series minimal models have modular invariant partition functions. Actually, the D-series minimal models too. 

\begin{exo}[Characters of minimal models]
 ~\label{exo:chmm}
 Show that the characters of fully degenerate representations in the Kac table of the $(p, q)$ minimal model are
 \begin{align}
 \chi_{\langle r,s\rangle}(\tau) = \sum_{k\in\mathbb{Z}} \left( \chi_{P_{\langle r,s\rangle} + ik\sqrt{pq}} - \chi_{P_{\langle r,-s\rangle} + ik\sqrt{pq}}\right)\ .
 \label{eq:chmm}
\end{align}
\end{exo}

\begin{exo}[Modular $S$-matrices of minimal models]
~\label{exo:mods}
 Show that the modular $S$-matrix of the $(p, q)$ minimal model is 
 \begin{align}
 S_{\langle r,s\rangle, \langle r', s'\rangle} = -\sqrt{\frac{8}{pq}}(-1)^{rs'+r's} \sin\left(\pi\frac{q}{p}rr'\right)\sin\left(\pi\frac{p}{q}ss'\right)\ .
 \label{eq:smm}
\end{align}
\end{exo}


\section{Solutions of Exercises}


\paragraph{Exercise \ref{exo:sphere}.}

Let us introduce the map
\begin{align}
 g = \left(\begin{array}{cc} a & b \\ c & d \end{array}\right) \in GL_2({\mathbb{C}}) \quad \longmapsto\quad f_g(z) = \frac{az+b}{cz+d}\ .
\end{align}
A direct calculation shows that $f_{g_1}\circ f_{g_2} = f_{g_1g_2}$, so that our map is a group morphism from $GL_2(\mathbb{C})$ to the global conformal group of the sphere. The morphism is manifestly surjective, let us determine its kernel. The matrix $g$ belongs to the kernel if and only if $\frac{az+b}{cz+d} =z$, equivalently $g=\left(\begin{array}{cc} a & 0 \\ 0 & a \end{array}\right)$ for $a\in\mathbb{C}^*$. So the global conformal group can be written as $\frac{GL_2(\mathbb{C})}{\mathbb{C}^*}$. Now, modulo $\mathbb{C}^*$, any element of $GL_2(\mathbb{C})$ is equivalent to a matrix of determinant one, i.e. an element of $SL_2(\mathbb{C})$. So the map $g\longmapsto f_g$ is also surjective as a map from $SL_2(\mathbb{C})$ to the global conformal group, and in $SL_2(\mathbb{C})$ its kernel is $\mathbb{Z}_2$, since $\det\left(\begin{array}{cc} a & 0 \\ 0 & a \end{array}\right)=1 \iff a\in\{1,-1\}$. Therefore, the global conformal group can also be written as $\frac{SL_2(\mathbb{C})}{\mathbb{Z}_2}$. 

\paragraph{Exercise \ref{exo:vir}.}

Let us look for central extensions of the Witt algebra, starting with the ansatz
\begin{align}
  [\mathbf 1, L_n] = 0 \quad , \quad [L_n,L_m] = (n-m)L_{n+m} +f(n,m)\mathbf 1 \ ,
  \label{eq:wext}
 \end{align}
for some function $f(n,m)$. The constraint on $f(n,m)$ from antisymmetry $[L_n,L_m]=-[L_m,L_n]$ is 
\begin{align}
 f(n,m)= - f(m,n)\ .
\end{align}
The constraint from the Jacobi identity $[[L_n,L_m],L_p] + [[L_m,L_p],L_n] + [[L_p,L_n],L_m]=0$ is 
\begin{align}
 (n-m)f(n+m,p) + (m-p)f(m+p,n) + (p-n)f(n+p,m) = 0 \ .
 \label{eq:jac}
\end{align}
In the case $p=0$, this reduces to
\begin{align}
 (m+n)f(n,m)+(m-n)f(m+n,0)=0\ .
 \label{eq:jac0}
\end{align}
This means that for $m+n\neq 0$, $f(n,m)$ can be written in terms of a function of only one variable. We can actually set this function to zero by reparametrizing the generators of our algebra. If indeed we define 
\begin{align}
 L_n = L'_n + g(n) \mathbf 1\ ,
\end{align}
for some function $g(n)$, then the generators $L'_n$ obey commutation relations of the type \eqref{eq:wext}, with however the function 
\begin{align}
 f'(n,m) = f(n, m)+(n-m)g(n+m)\ .
\end{align}
Let us choose $g(n) = -\frac{f(n,0)}{n}$ for $n\neq 0$, then eq. \eqref{eq:jac0} becomes $f'(n,m)=0$ for $n+m\neq 0$. We therefore write 
\begin{align}
 f'(n,m) = \delta_{n+m,0}h(n)\ , 
\end{align}
for some unknown function $h(n)$ such that $h(0)=0$ by antisymmetry. We still have the freedom to choose $g(0)$, with $f'(n,-n) = f(n,-n) + 2ng(0)$. We use this freedom for setting $h(1)=0$. Let us rewrite the Jacobi identity \eqref{eq:jac} in terms of the function $h(n)$:
\begin{align}
 (n-m)h(m+n) -(2m+n)h(n) +(m+2n)h(m) = 0\ .
 \label{eq:hhh}
\end{align}
In the particular case $m=1$, this becomes $(n-1)h(n+1) - (n+2)h(n)=0$. Since $h(-1)=h(0)=h(1)=0$ it is natural to write $h(n) = (n-1)n(n+1) h'(n)$, and our equation becomes $h'(n+1)=h'(n)$.  This shows that $h(n)=\lambda n(n^2-1)$ for some constant $\lambda$. To conclude, it remains to check that this solves eq. \eqref{eq:hhh} not only for $m=1$, but for all values of $m$ -- a straightforward computation.

\paragraph{Exercise \ref{exo:level2}.} 

Straighforward calculations.

\paragraph{Exercise \ref{exott}.} 

As a consequence of eq. \eqref{eq:lvtv} together with $T(y)T(z)=T(z)T(y)$, we have 
\begin{align}
 [L_n^{(z_0)},L_m^{(z_0)}] = -\frac{1}{4\pi^2} \left(\oint_{z_0} dy \oint_{z_0} dz - \oint_{z_0} dz \oint_{z_0} dy\right) (y-z_0)^{n+1}(z-z_0)^{m+1} T(y)T(z)\ ,
 \label{lzlz}
\end{align}
In this formula, $\oint_{z_0} dy \oint_{z_0} dz$ means that the integration over $z$ should be performed before the integration over $y$, so the contour of integration over $z$ should be inside the contour of integration over $y$. We have a second term where the positions of the contours are exchanged. Let us focus on the contribution of a given value of $z$: in the second term this is simply $\oint_{z_0} dy$, while in the first term this is $\oint_{z} dy + \oint_{z_0} dy$.
Therefore, we find
\begin{align}
 \oint_{z_0} dy \oint_{z_0} dz - \oint_{z_0} dz \oint_{z_0} dy = \oint_{z_0} dz \oint_z dy\ ,
\end{align}
and therefore,
\begin{align}
 [L_n^{(z_0)},L_m^{(z_0)}] = -\frac{1}{4\pi^2} \oint_{z_0} dz \oint_{z} dy\  (y-z_0)^{n+1}(z-z_0)^{m+1} T(y)T(z)\ .
\end{align}
Let us compute the integral over $y$, using the OPE \eqref{tt}. We find 
\begin{multline}
 \frac{1}{2\pi i} \oint_z dy\ (y-z_0)^{n+1} T(y)T(z) = \frac{c}{12} n(n^2-1) (z-z_0)^{n-2} 
 \\
 +2(n+1)(z-z_0)^n T(z) + (z-z_0)^{n+1} \partial T(z)\ .
 \label{eq:oitt}
\end{multline}
It remains to perform the integration over $z$. Using eq. \eqref{eq:lvtv}, this shows that $[L_n^{(z_0)},L_m^{(z_0)}]$ is given by the Virasoro algebra's commutation relations, with the last two terms of eq. \eqref{eq:oitt} contributing respectively $(2n+2)L_{m+n}^{(z_0)}$ and $-(m+n+2)L_{m+n}^{(z_0)}$.

\paragraph{Exercise \ref{exo:4pt}.} 

If $\Big<V_\Delta(z_1)V_\Delta(z_2)\Big> = (z_1-z_2)^{-2\Delta}$, then
\begin{align}
 \Big<V_\Delta(\infty)V_\Delta(z_2)\Big> = \lim_{z\to\infty} z^{2\Delta}(z-z_2)^{-2\Delta}=1 \ .
\end{align}
Similarly, we compute 
\begin{align}
 \Big<V_{\Delta_1}(\infty)V_{\Delta_2}(z_2)V_{\Delta_3}(z_3)\Big> \propto (z_2-z_3)^{\Delta_1-\Delta_2-\Delta_3}\ .
\end{align}
And the computation for the four-point function is straightforward.

Using eq. \eqref{eq:zgc} with $\left(\begin{smallmatrix} a & b \\ c & d \end{smallmatrix} \right) = \left(\begin{smallmatrix} 0 & -1 \\ 1 & 0 \end{smallmatrix}\right)$, we find 
\begin{align}
 \left< \prod_{i=1}^N  V_{\Delta_i}\left(-\tfrac{1}{z_i}\right) \right>
 = \prod_{i=1}^N z_i^{2\Delta_i} \left< \prod_{i=1}^N V_{\Delta_i}(z_i) \right>\ .
\end{align}
Since $V_{\Delta_1}(-\frac{1}{z_1})$ has a finite limit as $z_1\to \infty$, we deduce that $z_1^{2\Delta_1}V_{\Delta_1}(z_1)$ has a finite limit too, provided $z_2,\dots, z_N$ are finite.

\paragraph{Exercise \ref{exo:bpz3pt}.} 

Straightforward calculations lead to 
\begin{align}
 \left< V_{\langle 2, 1 \rangle} V_{\Delta_2} V_{\Delta_3} \right> \neq 0 \quad \implies \quad 
 2(\Delta_2-\Delta_3)^2 +b^2(\Delta_2+\Delta_3) -2\Delta_{\langle 2, 1 \rangle}^2 -b^2\Delta_{\langle 2, 1 \rangle} = 0\ .
 \end{align}
 It only remains to replace conformal dimensions with momentums.

\paragraph{Exercise \ref{exo:bpz}.}

The calculations using eq. \eqref{eq:4pt} are straightforward but tedious. Let us try to do a bit better. What prevents us from directly applying the BPZ equation to $G(z)=\Big< V_{\langle 2, 1 \rangle}(z) V_{\Delta_1}(0)V_{\Delta_2}(\infty)V_{\Delta_3}(1) \Big>$, is that the equation involves derivatives with respect to the three positions that we want to set to fixed values. So apparently we need to introduce $\Big< V_{\langle 2, 1 \rangle}(z) V_{\Delta_1}(z_1)V_{\Delta_2}(z_2)V_{\Delta_3}(z_3) \Big>$. 
However, using the global Ward identities, any derivative with respect to $z_1,z_2,z_3$ can be rewritten as a derivative with respect to $z$, and therefore eliminated from the BPZ equation. To do this efficiently, remember that the derivatives with respect to $z_i$ originated from using eq. \eqref{eq:ltv} for computing $\Big< L_{-2}V_{\langle 2, 1 \rangle}(z) V_{\Delta_1}(z_1)V_{\Delta_2}(z_2)V_{\Delta_3}(z_3) \Big>$. In order to eliminate such derivatives, it is enough to compute the following expression instead:
\begin{align}
 \frac{1}{2\pi i}\oint_{z} dy\frac{\prod_{i=1}^3(y-z_i)}{y-z} Z(y) \ .
\end{align}
Closing the contour on $y=z$ gives us a combination of $L_{-2}^{(z)}$, $\frac{\partial}{\partial z}$, and scalar factors, acting on  $\Big< V_{\langle 2, 1 \rangle}(z) V_{\Delta_1}(z_1)V_{\Delta_2}(z_2)V_{\Delta_3}(z_3) \Big>$. Closing the contour on $y=z_i$ instead does not produce any derivatives with respect to $z_i$, thanks to the vanishing of the prefactor $\frac{\prod_{i=1}^3(y-z_i)}{y-z}$ at $y=z_i$. This leads to a version of the BPZ equation that involves derivatives with respect to $z$ only:
\begin{multline}
  \left\{ \prod_{i=1}^3(z-z_i)\left(-\frac{1}{b^2}\frac{\partial^2}{\partial z^2} +\sum_{i=1}^3 \frac{1}{z-z_i} {\frac{\partial}{\partial z}} \right) + (3z-z_1-z_2-z_3)\Delta_{\langle 2,1 \rangle} \right.
  \\
\left.  +\frac{z_{12}z_{13}}{z_1-z}\Delta_1 + \frac{z_{21}z_{23}}{z_2-z}\Delta_2+\frac{z_{31}z_{32}}{z_3-z}\Delta_3\right\} 
\left\langle V_{\langle 2,1 \rangle}(z)\prod_{i=1}^3 V_{\Delta_i}(z_i)\right\rangle  = 0\ .
\label{uode}
\end{multline}
In this version of the BPZ equation, it is straightforward to send $z_1,z_2,z_3$ to $0,\infty, 1$, and we obtain eq. \eqref{eq:ode}.

\paragraph{Exercise \ref{exo:ope}.}

Let us insert $\oint_C dz(z-z_2)^2 T(z)$ on both sides of eq. \eqref{eq:ope}, for $C$ a contour around both $z_1$ and $z_2$. Neglecting the dependence on $\bar z_i$, we rewrite this OPE as 
\begin{align}
 V_{\Delta_1}(z_1) V_{\Delta_2}(z_2) 
 = \sum_{\Delta\in S} C_{\Delta_1,\Delta_2,\Delta} z_{12}^{\Delta-\Delta_1-\Delta_2}
 \Big(V_{\Delta}(z_2) + f z_{12} L_{-1}V_{\Delta}(z_2) + O(z_{12}^2) \Big)\ .
\end{align}
We first compute the left-hand side. The integrand $(z-z_2)^2 T(z) V_{\Delta_1}(z_1)V_{\Delta_2}(z_2)$ is regular at $z=z_2$, and we only pick contributions from the pole at $z=z_1$. Using the OPE $T(z)V_{\Delta_1}(z_1)$ \eqref{eq:tvd}, we find 
\begin{align}
 \frac{1}{2\pi i}\oint_C dz(z-z_2)^2 T(z)V_{\Delta_1}(z_1)V_{\Delta_2}(z_2) = \left(2z_{12}\Delta_1+z_{12}^2\frac{\partial}{\partial z_1}\right) V_{\Delta_1}(z_1) V_{\Delta_2}(z_2)\ .
\end{align}
Using the OPE \eqref{eq:ope}, we compute 
\begin{multline}
 \frac{1}{2\pi i}\oint_C dz(z-z_2)^2 T(z)V_{\Delta_1}(z_1)V_{\Delta_2}(z_2)
 \\
 = \sum_{\Delta\in S} C_{\Delta_1,\Delta_2,\Delta} z_{12}^{\Delta-\Delta_1-\Delta_2+1} 
 \Big( (\Delta+\Delta_1-\Delta_2)V_{\Delta}(z_2) + O(z_{12}) \Big)\ .
 \label{eq:lhsope}
\end{multline}
Now we insert $\oint_C dz(z-z_2)^2 T(z)$ on the right-hand side of eq. \eqref{eq:ope}For any field $V(z_2)$ (primary or descendent) we have $\frac{1}{2\pi i}\oint_C dz(z-z_2)^2 T(z)V(z_2) = L_1 V(z_2)$. Since $L_1 V_{\Delta_2}(z_2)=0$, the leading contribution is from the level one descendent $L_{-1} V_{\Delta_2}(z_2)$,
\begin{multline}
 \frac{1}{2\pi i}\oint_C dz(z-z_2)^2 T(z)V_{\Delta_1}(z_1)V_{\Delta_2}(z_2)
 \\
 = \sum_{\Delta\in S} C_{\Delta_1,\Delta_2,\Delta} z_{12}^{\Delta-\Delta_1-\Delta_2+1} 
 \Big( f L_1L_{-1}V_{\Delta}(z_2) + O(z_{12}) \Big)\ .
\end{multline}
Using $L_1L_{-1}V_{\Delta}(z_2)=2\Delta V_\Delta(z_2)$, and comparing with the left-hand side result \eqref{eq:lhsope}, this leads to 
\begin{align}
 f = \frac{\Delta+\Delta_1-\Delta_2}{2\Delta}\ .
\end{align}
Of course, this expression is also the coefficient of the right-moving descendent $\bar L_{-1} V_\Delta(z_2)$.

\paragraph{Exercise \ref{exo:id}.} 

Let us write the OPE \eqref{eq:ope} in the case of the degenerate field $V_{\langle 1,1\rangle}$, while omitting the dependence on $\bar z_i$:
\begin{align}
 V_{\langle 1,1\rangle}(z_1) V_{\Delta_2}(z_2) = \sum_{\Delta\in S} C_{\langle 1,1\rangle,\Delta_2,\Delta}  \sum_{i=0}^\infty z_{12}^{\Delta-\Delta_2+i} \mathcal{L}_i V_\Delta(z_2) \ ,
\end{align}
where $\mathcal{L}_i V_\Delta(z_2)$ is some descendent at level $i$, and $\mathcal{L}_0 V_\Delta(z_2) = V_\Delta(z_2)$. Since $\frac{\partial}{\partial z_1} V_{\langle 1,1\rangle}(z_1) =0 $, we have 
\begin{align}
 0 = \sum_{\Delta\in S} C_{\langle 1,1\rangle,\Delta_2,\Delta}  \sum_{i=0}^\infty z_{12}^{\Delta-\Delta_2+i-1}(\Delta-\Delta_2+i) \mathcal{L}_i V_\Delta(z_2) \ .
\end{align}
Assuming $C_{\langle 1,1\rangle,\Delta_2,\Delta}\neq 0$, the vanishing of the leading $i=0$ term implies $\Delta_2=\Delta$. The vanishing of an $i>0$ term then implies $\mathcal{L}_i V_\Delta(z_2) =0$. Therefore, the OPE reduces to 
\begin{align}
 V_{\langle 1,1\rangle}(z_1) V_{\Delta}(z_2) = C_\Delta V_\Delta(z_2)\ ,
\end{align}
where $C_\Delta = C_{\langle 1,1\rangle,\Delta,\Delta} $. Let us use this OPE in a correlation function that involves the fields $V_{\langle 1,1\rangle}(z_1)V_{\Delta_2}(z_2) V_{\Delta_3}(z_3)$. Using commutativity and associativity of the OPE, we obtain
\begin{align}
 V_{\langle 1,1\rangle}(z_1)V_{\Delta_2}(z_2) V_{\Delta_3}(z_3) = C_{\Delta_2} V_{\Delta_2}(z_2) V_{\Delta_3}(z_3) = C_{\Delta_3} V_{\Delta_2}(z_2) V_{\Delta_3}(z_3)\ .
\end{align}
This implies $C_{\Delta_2}=C_{\Delta_3}$, and actually $C_\Delta$ cannot depend on $\Delta$.

\paragraph{Exercise \ref{exo:block}.}

Using the OPE \eqref{eq:ope} including the first subleading correction from Exercise \ref{exo:ope}, and omitting the dependence on $\bar z_i$, we have 
\begin{multline}
 \Big<V_{\Delta_1}(z)V_{\Delta_2}(0)V_{\Delta_3}(\infty)V_{\Delta_4}(1)\Big> = \sum_{\Delta\in S} C_{\Delta_1,\Delta_2,\Delta} z^{\Delta-\Delta_1-\Delta_2}
 \\ \times 
 \left< \left(1 + \frac{\Delta+\Delta_1-\Delta_2}{2\Delta}z L_{-1} + O(z^2)\right) V_\Delta(0) V_{\Delta_3}(\infty)V_{\Delta_4}(1)\right>\ .
\end{multline}
Let us then compute $\Big< L_{-1} V_\Delta(0) V_{\Delta_3}(\infty)V_{\Delta_4}(1)\Big>$. This is done by computing it for three arbitrary field positions before specializing to $0,1,\infty$, using $L_{-1} V_\Delta(z) =\frac{\partial}{\partial z} V_\Delta(z)$. The result is 
\begin{align}
 \Big< L_{-1} V_\Delta(0) V_{\Delta_3}(\infty)V_{\Delta_4}(1)\Big> = C_{\Delta,\Delta_3,\Delta_4} (\Delta+\Delta_4-\Delta_3)\ .
\end{align}
We therefore deduce
\begin{multline}
 \Big<V_{\Delta_1}(z)V_{\Delta_2}(0)V_{\Delta_3}(\infty)V_{\Delta_4}(1)\Big> = \sum_{\Delta\in S} C_{\Delta_1,\Delta_2,\Delta} C_{\Delta,\Delta_3,\Delta_4} z^{\Delta-\Delta_1-\Delta_2}
 \\ \times 
 \left( 1+ \frac{(\Delta+\Delta_1-\Delta_2)(\Delta+\Delta_4-\Delta_3)}{2\Delta}z + O(z^2)\right) \ ,
\end{multline}
which is equivalent to eq. \eqref{eq:fsl}. The first subleading term has a pole at $\Delta=0$, with the residue $(\Delta_1-\Delta_2)(\Delta_4-\Delta_3)$. The residue vanishes if $\Delta_1=\Delta_2$ or $\Delta_3=\Delta_4$, i.e. if at least one of the three-point functions $\left< V_{\langle 1,1\rangle} V_{\Delta_1}V_{\Delta_2}\right>$ and $\left< V_{\langle 1,1\rangle} V_{\Delta_3}V_{\Delta_4}\right>$ is non-vanishing.

\paragraph{Exercise \ref{exo:hdr}.} 

We already know that the fusion product \eqref{rtv} holds for the representations $\mathcal{R}_{\langle 1,1\rangle}$, $\mathcal{R}_{\langle 2,1\rangle}$ and $\mathcal{R}_{\langle 1,2\rangle}$. Let us prove it more generally by recursion on $r,s$. Let us assume that it holds for all $r\leq r_0$ and $s\leq s_0$. For $s\leq s_0$ we compute 
\begin{align}
 \mathcal{R}_{\langle 2,1\rangle} \times \mathcal{R}_{\langle r_0, s\rangle} \times \mathcal{V}_P 
 &= \sum_{i=-\frac{r_0-1}{2}}^{\frac{r_0-1}{2}} 
 \sum_{j=-\frac{s-1}{2}}^{\frac{s-1}{2}}  
 \mathcal{R}_{\langle 2,1\rangle} \times \mathcal{V}_{P + ib+jb^{-1}} \ ,
 \\
 &= \left\{\sum_{i=-\frac{r_0}{2}}^{\frac{r_0}{2}} + \sum_{i=-\frac{r_0-2}{2}}^{\frac{r_0-2}{2}} \right\}
 \sum_{j=-\frac{s-1}{2}}^{\frac{s-1}{2}} \mathcal{V}_{P + ib+jb^{-1}} \ .
 \\
 &= \mathcal{R}_{\langle r_0-1,s\rangle} \times \mathcal{V}_P + \sum_{i=-\frac{r_0}{2}}^{\frac{r_0}{2}}\sum_{j=-\frac{s-1}{2}}^{\frac{s-1}{2}} \mathcal{V}_{P + ib+jb^{-1}} \ .
\end{align}
(This is true even if $r_0=1$, with the convention $\mathcal{R}_{\langle 0,s\rangle}=0$.) We obtain a combination of finitely many Verma modules, which shows that $\mathcal{R}_{\langle 2,1\rangle} \times \mathcal{R}_{r_0, s} $ must be a degenerate representation. In this degenerate representation, we know that the highest-weight states have the momentums $P_{\langle r_0, s\rangle} \pm \frac{b}{2} = P_{\langle r_0\pm 1,s\rangle}$. We already know that the momentum $P_{\langle r_0-1,s\rangle}$ corresponds to the degenerate representation $\mathcal{R}_{\langle r_0-1,s\rangle}$, and there must be a degenerate representation with the momentum  $P_{\langle r_0+1,s\rangle}$, which we call $\mathcal{R}_{\langle r_0+1,s\rangle}$. We can similarly do the recursion on $s$, and we obtain degenerate representations $\mathcal{R}_{\langle r,s\rangle}$ with $r,s\in\mathbb{N}^*$.

The same method can be used for determining the fusion products \eqref{rrsr} of degenerate representations by recursion on $r_1,s_1$, starting with the known cases $(r_1,s_1)\in \{(1,1), (2,1), (1,2)\}$.

\paragraph{Exercise \ref{exo:cmm}.} 

Let $\mathcal{R}_{\langle r_1,s_1\rangle}$ and $\mathcal{R}_{\langle r_2,s_2\rangle}$ be two doubly degenerate representations in $S_{p, q}$. The fusion product $\mathcal{R}_{\langle r_1,s_1\rangle}\times \mathcal{R}_{\langle r_2,s_2\rangle}$ must be a subspace of two fusion products computed with the rule \eqref{rrsr} for degenerate representations: $\mathcal{R}_{\langle r_1,s_1\rangle}\times \mathcal{R}_{\langle r_2,s_2\rangle}$ and $\mathcal{R}_{\langle r_1,s_1\rangle}\times \mathcal{R}_{\langle p-r_2,q-s_2\rangle}$. The determination of this subspace is not completely straightforward, because each representation that appears in one of our two fusion products can be written in two possible ways, $\mathcal{R}_{\langle r_3, s_3\rangle}$ or $\mathcal{R}_{\langle p-r_3, q-s_3\rangle}$. 
This ambiguity can be lifted by writing the two fusion products such that the parities of $r_3,s_3$ are the same in one fusion products as in the other. (Remember that at least one of the integers $p,q$ must be odd.)
In particular, if we write $\mathcal{R}_{\langle r_1,s_1\rangle}\times \mathcal{R}_{\langle r_2,s_2\rangle}$ as in eq. \eqref{rrsr}, we should not write $\mathcal{R}_{\langle r_1,s_1\rangle}\times \mathcal{R}_{\langle p-r_2,q-s_2\rangle}$ as 
\begin{align}
 \mathcal{R}_{\langle r_1,s_1\rangle}\times \mathcal{R}_{\langle p-r_2,q-s_2\rangle} 
 = \sum_{r_3\overset{2}{=}|r_1+r_2-p|+1}^{r_1-r_2+p-1}\ \sum_{s_3\overset{2}{=}|s_1+s_2-q|+1}^{s_1-s_2+q-1} \mathcal{R}_{\langle r_3,s_3 \rangle}\ ,
\end{align}
but rather as 
\begin{align}
  \mathcal{R}_{\langle r_1,s_1\rangle}\times \mathcal{R}_{\langle p-r_2,q-s_2\rangle} 
  = \sum_{r_3\overset{2}{=}r_2-r_1+1}^{\min(r_1+r_2,2p-r_1-r_2)-1}\ \sum_{s_3\overset{2}{=}s_2-s_1+1}^{\min(s_1+s_2,2q-s_1-s_2)-1} \mathcal{R}_{\langle r_3,s_3 \rangle}\ .
\end{align}
In this form, we see that the intersection of the two fusion products $\mathcal{R}_{\langle r_1,s_1\rangle}\times \mathcal{R}_{\langle r_2,s_2\rangle}$ and $\mathcal{R}_{\langle r_1,s_1\rangle}\times \mathcal{R}_{\langle p-r_2,q-s_2\rangle}$ of degenerate representations is given by eq. \eqref{rrmm}. Taking further intersections with $\mathcal{R}_{\langle p-r_1,q-s_1\rangle}\times \mathcal{R}_{\langle r_2,s_2\rangle}$ and $\mathcal{R}_{\langle p-r_1,q-s_1\rangle}\times \mathcal{R}_{\langle p-r_2,q-s_2\rangle}$ does not yield further constraints, and the fusion products of our doubly degenerate representations are therefore given by eq. \eqref{rrmm}.

\paragraph{Exercise \ref{exo:frd}.}

Fusion rules are a priori defined for representations of the Virasoro algebra, and do not know how the left- and right-moving representations are combined. Therefore, fusion rules may not unambiguously determine operator product expansions in models where the same left-moving Virasoro representation is combined with different right-moving representations. 

However, assuming that $p$ is even, let us look at the values of the $r$ index in the D-series spectrum \eqref{eq:sds}. The parity of $r$ is unambiguously defined, as it does not change under $(r,s)\to (p-r,q-s)$. If moreover $p\equiv 0 \bmod 4$, then $r$ takes odd values in the diagonal sector, and even values in the non-diagonal sector: therefore, the two sectors do not involve the same representations of the Virasoro algebra. From the fusion product \eqref{rrmm}, we deduce that the fusion product of two representations is diagonal if and only if the two representations belong to the same sector. Let us write this in terms of OPEs of the primary fields $V^\epsilon_{\langle r,s\rangle}$ of the model, where the boolean $\epsilon = r-1\bmod 2$ indicates the sector: $V^0_{\langle r,s\rangle}$ corresponds to the representation $\mathcal{R}_{ \langle r,s\rangle} \otimes \bar{\mathcal{R}}_{\langle r,s\rangle}$, and $V^1_{\langle r,s\rangle}$ to $\mathcal{R}_{\langle r,s\rangle} \otimes \bar{\mathcal{R}}_{\langle p-r,s\rangle}$. We then have OPEs of the type 
\begin{align}
 V^{\epsilon_1}_{\langle r_1,s_1\rangle}V^{\epsilon_2}_{\langle r_2,s_2\rangle} \sim \sum_{r_3\overset{2}{=}|r_1-r_2|+1}^{\min(r_1+r_2,2p-r_1-r_2)-1}\ \sum_{s_3\overset{2}{=}|s_1-s_2|+1}^{\min(s_1+s_2,2q-s_1-s_2)-1} V^{\epsilon_1+\epsilon_2}_{\langle r_3,s_3\rangle}\ .
\end{align}
For $p\equiv 2\bmod 4$, we still use the boolean $\epsilon$ for labelling sectors, although it is no longer related to the parity of $r$, as $r$ is odd in both sectors. The same expression for the fusion rules makes sense, although the conservation of diagonality is now an ansatz to be tested, rather than an unavoidable consequence of symmetry.

\paragraph{Exercise \ref{exo:upsilon}.}

From eq. \eqref{eq:up}, we deduce how the function $\Upsilon_b(x)$ behaves under shifts:
\begin{align}
 \frac{\Upsilon_b(x+b)}{\Upsilon_b(x)} = \lambda_b^{2bx-1} \prod_{n=0}^\infty \left[ \frac{n+1-bx}{n+bx} e^{ \sum_{m=0}^\infty \frac{2bx-1}{\left(\frac{Q}{2}+mb+nb^{-1}\right)^2}}\right] \ .
\end{align}
From Weierstrass's definition of the Gamma function, we deduce the product formula 
\begin{align}
 \gamma(x) = e^{-\gamma_\text{EM}(2x-1)} \frac{1-x}{x} \prod_{n=1}^\infty \left[\frac{n+1-x}{n+x} e^{\frac{2x-1}{n}}\right] \ ,
\end{align}
where $\gamma_\text{EM}$ is the Euler--Mascheroni constant. This implies 
\begin{align}
 \frac{\Upsilon_b(x+b)}{\Upsilon_b(x)} =  \gamma(bx) e^{(2bx-1)(\log \lambda_b +\gamma_\text{EM}  + K_b)}\ ,
\end{align}
where we define 
\begin{align}
 K_b = \sum_{n=0}^\infty \left[\sum_{m=0}^\infty \frac{1}{\left(\frac{Q}{2}+mb+nb^{-1}\right)^2} - \frac{1}{n}\right] \ ,
\end{align}
with the convention $\frac{1}{0} = 0$. We are not there yet, because $K_b$ is not invariant under $b\to \frac{1}{b}$. To understand its behaviour under $b\to \frac{1}{b}$, we introduce a cutoff $M$ on the sum over $M$, and perform the split $\sum_{n=0}^\infty = \sum_{n=0}^M + \sum_{n=M+1}^\infty$. This leads to 
\begin{align}
 K_b - K_{b^{-1}} = \lim_{M\to \infty} \left[ \sum_{n=0}^M \sum_{m=M+1}^\infty - \sum_{m=0}^M \sum_{n=M+1}^\infty\right] \frac{1}{\left(\frac{Q}{2}+mb+nb^{-1}\right)^2}\ .
\end{align}
In order to compute this limit, we can replace the sums with integrals, and we obtain
\begin{align}
  K_b - K_{b^{-1}} = \lim_{M\to \infty} \left[\int_0^{Mb^{-1}} dy \int_{Mb}^\infty dx - \int_0^{Mb} dx\int_{Mb^{-1}}^\infty dy\right] \frac{1}{(x+y)^2} = -2\log b\ .
\end{align}
This shows that $K_b+\log b$ is invariant under $b\to \frac{1}{b}$. Therefore, we set 
\begin{align}
 \lambda_b = e^{-\gamma_\text{EM}-K_b-\log b}\ ,
\end{align}
and we obtain the shift equation \eqref{upup} for $\frac{\Upsilon_b(x+b)}{\Upsilon_b(x)}$. Since $\lambda_b$ is invariant under $b\to \frac{1}{b}$, the shift equation for  $\frac{\Upsilon_b(x+\frac{1}{b})}{\Upsilon_b(x)}$ follows immediately.

\paragraph{Exercise \ref{exo:fbs}.}

For $c=1$ the conformal dimension is related to the momentum by $\Delta = -\alpha^2$. If we have a state with momentums $(\alpha,\bar\alpha)$, its spin is $\bar\alpha^2-\alpha^2\in\mathbb{Z}$. Since we also assume that the spectrum is closed under fusion, and that we have a diagonal state with momentums $(\frac{i}{2R},\frac{i}{2R})$, we must have a state with momentums $(\alpha +\frac{i}{2R},\bar \alpha+\frac{i}{2R})$. The condition for this state to have integer spin is 
\begin{align}
 \left(\bar \alpha +\tfrac{i}{2R}\right)^2 - \left(\alpha +\tfrac{i}{2R}\right)^2 \in \mathbb{Z} \iff \frac{i}{R}(\alpha -\bar \alpha)\in\mathbb{Z}\ .
 \label{eq:lba}
\end{align}
We assume that there are nondiagonal states $(\alpha,\bar\alpha)$ that minimize this number, i.e. $\alpha-\bar\alpha = -iR$.
Since we also have $\bar\alpha^2-\alpha^2 = -(\alpha-\bar \alpha)(\alpha+\bar \alpha)\in\mathbb{Z}$, we deduce $\alpha+\bar\alpha\in \frac{i}{R}\mathbb{Z}$. Again, we assume that all allowed values actually occur, so that there is a state with $\alpha+\bar\alpha = 0$, i.e. a state with momentums $(-\frac{iR}{2},\frac{iR}{2})$.

For $(\alpha,\bar\alpha)$ any state, the condition for the spin of $(\alpha-\frac{iR}{2},\bar\alpha +\frac{iR}{2})$ to be integer is $iR(\alpha+\bar\alpha)\in\mathbb{Z}$. Adding the constraint \eqref{eq:lba}, we deduce that our state must belong to the spectrum $S_R$ \eqref{sr}.

For an arbitrary central charge, the conformal dimension is $\Delta = \alpha(Q-\alpha)$, and condition for  a state in the spectrum $S_R$ to have integer spin is 
\begin{align}
 Q(\alpha-\bar \alpha) \in\mathbb{Z} \iff QiRw \in\mathbb{Z}\ .
\end{align}
This must hold for any $w\in\mathbb{Z}$, and we must have $R\in \frac{i}{Q}\mathbb{Z}$. Therefore, compactified free bosons exist only for certain discrete values of the radius.

\paragraph{Exercise \ref{exo:chmm}.}

To do this exercise, we only need to know the identities 
\begin{align}
 \Delta_{\langle r,s\rangle} = \Delta_{\langle -r,-s\rangle}=\Delta_{\langle p+r,q+s\rangle}\ ,
 \label{eq:deltaid}
\end{align}
and the fact that a primary state with dimension $\Delta_{\langle r,s\rangle}$ with $r,s\in\mathbb{N}^*$ has a null vector with dimension $\Delta_{\langle r,-s\rangle}$.

By definition, the doubly degenerate representation $\mathcal{R}_{\langle r,s\rangle} =\mathcal{R}_{\langle p-r,q-s\rangle}$ has two vanishing null vectors, with the dimensions $\Delta_{\langle r, -s\rangle}$ and $\Delta_{\langle p-r,s-q\rangle}$. Actually it has infinitely many null vectors due to the identities \eqref{eq:deltaid}, but we only consider the two null vectors with the lowest conformal dimensions for the moment. 
The null vector with dimension $\Delta_{\langle r, -s\rangle}=\Delta_{\langle p+r, q-s\rangle} = \Delta_{\langle p-r,q+s\rangle}$ itself has null vectors with dimensions $\Delta_{\langle p+r, s-q\rangle}$ and $\Delta_{\langle p-r,-q-s\rangle}$. One way to tame this proliferation of null vectors is to use the identities \eqref{eq:deltaid} in order to rewrite them all with the same first index $r$, and to label them by their second index. In this notation, the primary state $s$ has null vectors $-s$ and $2q-s$. The first one of these null vectors itself has null vectors $s-2q$ and $2q+s$. Interestingly, the second null vector has two null vectors with the very same labels $s-2q$ and $2q+s$. We assume that two null vectors with the same conformal dimension are actually identical: this means that the submodules generated by the null vectors $-s$ and $2q-s$ share two submodules. The character of our representation therefore starts with 
\begin{align}
 \chi_{\langle r,s\rangle} = \chi_s - \chi_{-s} -\chi_{-s+2q} + \chi_{s-2q} + \chi_{s+2q} + \cdots \ , 
\end{align}
where $\chi_s$ is a temporary notation for the Verma module character $\chi_{P_{\langle r,s\rangle}}$, and the two shared submodules had to be added back. Iterating the reasoning, we find 
\begin{align}
 \chi_{\langle r,s\rangle} = \sum_{k\in\mathbb{Z}} \left(\chi_{s + 2k q} -\chi_{-s+2k q}\right)\ . 
\end{align}
Using $P_{\langle r,s+2q\rangle} = P_{\langle r,s\rangle} + i\sqrt{pq}$, this can be rewritten as eq. \eqref{eq:chmm}.

But what do we do with the infinitely many null vectors that we have neglected? Actually, nothing: the neglected null vectors all coincide with null vectors that we already did take into account, and do not affect the character.

\paragraph{Exercise \ref{exo:mods}.}

From the Fourier transforms of Gaussian functions, we deduce the modular transformation of the character $\chi_{P}(\tau)$ of a Verma module: 
 \begin{align}
  \chi_P(\tau) = \sqrt{2}i\int_{i\mathbb{R}} dP'\ e^{4\pi iPP'} \chi_{P'}(-\tfrac{1}{\tau})\ . 
 \end{align}
Using the Poisson resummation formula $\sum_{k\in\mathbb{Z}} e^{2\pi ikx} = \sum_{\ell\in\mathbb{Z}} \delta(x+\ell)$, we then deduce the modular transformation of the characters of fully degenerate representation in the Kac table of the $(p, q)$ minimal model,
\begin{align}
 \chi_{\langle r,s\rangle}(\tau) = 
 -\sqrt{\frac{2}{pq}}\sum_{\ell\in\mathbb{Z}} 
 \sin\left(\pi\frac{r}{p}\ell\right) 
 \sin\left(\pi\frac{s}{q}\ell\right)
 \chi_{i\frac{\ell}{2\sqrt{pq}}}(-\tfrac{1}{\tau})\ .
 \label{eq:chs}
\end{align}
Let us insert $\sum_{r=1}^{p-1} (-1)^{rs'}\sin\left(\pi\frac{q}{p}rr'\right)$ on both sides of this equation. We also insert an analogous sum over $s$. The right-hand side is computed using the elementary identity
\begin{align}
 \sum_{r=1}^{p-1} (-1)^{rs'}\sin\left(\pi\frac{q}{p}rr'\right)\sin\left(\pi\frac{r}{p}\ell\right) 
 = -\frac{p}{2}\sum_{\epsilon =\pm} \epsilon \delta_{\ell\equiv\epsilon qr'+ps'\bmod 2p}\ ,
\end{align}
and it can then be written in terms of characters of fully degenerate representations, leading to 
eq. \eqref{eq:smm}.


\hypersetup{linkcolor=black}

\renewcommand{\listtheoremname}{List of Exercises}
\listoftheorems[ignoreall, show={exo}]

\hypersetup{linkcolor=blue}


\end{document}